\documentclass[sigconf]{acmart}

\usepackage{soul}
\usepackage{booktabs} % For formal tables
\usepackage{enumitem}
\usepackage{boldline,multirow}
\usepackage{array}
\usepackage{graphicx}
\usepackage{pdfpages}
\usepackage{algorithm}
\usepackage{algorithmic}
\usepackage{diagbox}
\usepackage{url}
\usepackage{flushend}
\usepackage{xcolor}
\usepackage{float}
\acmDOI{10.475/123_4}

\acmISBN{123-4567-24-567/08/06}

\copyrightyear{2019}
\acmYear{2019}
\acmConference[CIKM '19]{The 28th ACM International Conference on Information and Knowledge Management}{November 3--7, 2019}{Beijing, China}
\acmBooktitle{The 28th ACM International Conference on Information and Knowledge Management (CIKM '19), November 3--7, 2019, Beijing, China}
\acmPrice{15.00}
\acmDOI{10.1145/3357384.3357979}
\acmISBN{978-1-4503-6976-3/19/11}

\begin{document}
\fancyhead{}
\title[\model]{Learning to Identify High Betweenness Centrality Nodes from Scratch: A Novel Graph Neural Network Approach}
\titlenote{This work was done when the first author was a visiting student at UCLA.}

\author[C. Fan, L. Zeng, Y. Ding, M. Chen, Y. Sun, Z. Liu]{Changjun Fan$^{1, 2}$, Li Zeng$^1$, Yuhui Ding$^3$, Muhao Chen$^{2, 4}$, Yizhou Sun$^2$, Zhong Liu$^1$}

\affiliation{%
  \institution{$^1$College of Systems Engineering, National University of Defense Technology}
  \institution{$^2$Department of Computer Science, University of California, Los Angeles}
  \institution{$^3$Department of Computer Science and Technology, Tsinghua University}
  \institution{$^4$Department of Computer and Information Science, University of Pennsylvania}
  }

\email{{cjfan2017, muhaochen, yzsun}@ucla.edu, {zlli, liuzhong}@nudt.edu.cn, dingyh15@mails.tsinghua.edu.cn} 

\begin{comment}

\end{comment}

\newcommand{\model}{DrBC}

\newcommand{\stitle}[1]{\vspace{0.3ex}\noindent{\bf #1}}
\newcommand{\todo}[1]{{\color{red}\textbf{#1}}}
\def\st#1{~}
\newcommand{\dsnamens}{DATASET}
\newcommand{\dsname}{{\dsnamens}\ }
\newcommand{\ggnamens}{\emph{our corporation (anonymized for blind review)}}
\newcommand{\ggname}{{\ggnamens}\ }
\def\bhline{\specialrule{.2em}{0em}{0em}}
\newcolumntype{?}{!{\vrule width .15em}}

\newcommand{\fixme}[2][]{\todo[color=yellow,size=\scriptsize,fancyline,caption={},#1]{#2}} % to mark stuff that you know is missing or wrong when you write the text
\newcommand{\note}[4][]{\todo[author=#2,color=#3,size=\scriptsize,fancyline,caption={},#1]{#4}} % default note settings, used by macros below.

\newcolumntype{I}{!{\vrule width 1.5pt}}
\newlength\savedwidth
\newcommand\whline{\noalign{\global\savedwidth\arrayrulewidth
                            \global\arrayrulewidth 3pt}%
                   \hline
                   \noalign{\global\arrayrulewidth\savedwidth}}
\newlength\savewidth
\newcommand\shline{\noalign{\global\savewidth\arrayrulewidth
                            \global\arrayrulewidth 1.5pt}%
                   \hline
                   \noalign{\global\arrayrulewidth\savewidth}}

\newcommand{\nop}[1]{}
\newcommand{\YS}[1]{{\bf\color{red}[{\sc YS:} #1]}}

\newcommand{\muhao}[1]{{\color{blue}MC:\textbf{#1}}}
\newcommand{\yuhui}[1]{{\color{red}YH:\textbf{#1}}}
\newcommand{\FC}[1]{{\bf\color{blue}[{\sc FC:} #1]}}

\begin{abstract}
Betweenness centrality (BC) is a widely used centrality measures for network analysis, which seeks to describe the importance of nodes in a network in terms of the fraction of shortest paths that pass through them. It is key to many valuable applications, including community detection and network dismantling. Computing BC scores on large networks is computationally challenging due to its high time complexity. Many sampling-based approximation algorithms have been proposed to speed up the estimation of BC. However, these methods still need considerable long running time on large-scale networks, and their results are sensitive to even small perturbation to the networks.

In this paper, we focus on the efficient identification of top-$k$ nodes with highest BC in a graph, which is an essential task to many network applications. Different from previous heuristic methods, we turn this task into a learning problem and design an encoder-decoder based framework as a solution. Specifically, the encoder leverages the network structure to represent each node as an embedding vector, which captures the important structural information of the node. The decoder transforms each embedding vector into a scalar, which identifies the relative rank of a node in terms of its BC. We use the pairwise ranking loss to train the model to identify the orders of nodes regarding their BC. By training on small-scale networks, the model is capable of assigning relative BC scores to nodes for much larger networks, and thus identifying the highly-ranked nodes. Experiments on both synthetic and real-world networks demonstrate that, compared to existing baselines, our model drastically speeds up the prediction without noticeable sacrifice in accuracy, and even outperforms the state-of-the-arts in terms of accuracy on several large real-world networks.
\end{abstract}

%This context information will include user preferences for services and networks depending on different factors such as location, time, nearby people, battery resources etc, information about user schedules/agendas, activity e.g. running, walking, driving a car etc and past usage behavior.

\begin{CCSXML}
<ccs2012>
<concept>
<concept_id>10002950.10003624.10003633.10010917</concept_id>
<concept_desc>Mathematics of computing~Graph algorithms</concept_desc>
<concept_significance>500</concept_significance>
</concept>
<concept>
<concept_id>10003120.10003130.10003134.10003293</concept_id>
<concept_desc>Human-centered computing~Social network analysis</concept_desc>
<concept_significance>500</concept_significance>
</concept>
<concept>
<concept_id>10010147.10010257.10010293.10010294</concept_id>
<concept_desc>Computing methodologies~Neural networks</concept_desc>
<concept_significance>500</concept_significance>
</concept>
</ccs2012>
\end{CCSXML}

\ccsdesc[500]{Mathematics of computing~Graph algorithms}
\ccsdesc[300]{Computing methodologies~Neural networks}
%\ccsdesc[100]{Human-centered computing~Social network analysis}

\keywords{Betweenness Centrality, Learning-to-rank, Graph Neural Network} 

\maketitle

%\vspace{-0.5em}
\section{Introduction}
Betweenness centrality (BC) is a %very 
fundamental %measure
metric in the field of network analysis. %which describes the %effectivenness
It measures the significance of nodes %in connecting node pairs
in terms of their connectivity to other nodes via the shortest paths~\cite{mahmoody2016scalable}. %the network structure.
Numerous applications rely on the computation of BC, including community detection~\cite{newman2006modularity}, %identifying potential leaders
%leader identification in terrorist networks~\cite{krebs2002mapping}, routing optimization~\cite{dolev2010routing} 
network dismantling~\cite{braunstein2016network}, etc. 
The best known algorithm for computing %betweenness centrality
BC exactly is the Brandes algorithm~\cite{brandes2001faster}, %and its
whose time complexity is $O(|V||E|)$ on unweighted networks and $O(|V||E|+|V|^2log|V|)$ on weighted networks, respectively, where $|V|$ and $|E|$ denote the numbers of nodes and edges in the network. %Algorithms with such time complexity is prohibitive for large real networks~\cite{alghamdi2017benchmark}. 
%Take com-LiveJournal~\cite{yang2015defining} as an example, which is a large social network with 4M nodes and 34M edges, it takes Brandes algorithm more than a year to compute BC on a typical workstation \cite{alghamdi2017benchmark}. 
%Moreover, there are many cases requiring BC to be dynamically maintained~\cite{newman2006modularity,holme2002attack}, where the network structure keeps changing. The computation may not be able to finish before the network structure changes again.
Recently, extensive efforts have been made %on
to approximation algorithms for BC~\cite{riondato2016fast,mahmoody2016scalable,yoshida2014almost}. However, the accuracy of these algorithms decreases and the execution time increases considerably %with
along with the increase in the network size. Moreover, there are many cases requiring BC to be dynamically maintained~\cite{newman2006modularity}, where the network topology keeps changing. 
In large-scale online systems such as social networks and p2p networks,
the computation of BC may not finish before the network topology changes again.
%\muhao{There might be one or two sentences to bring out that identifying top K is an alternative but not well-addressed solution.}

In this paper, we %are interested in
focus on identifying nodes with high BC in large networks.
%based on the following observations. First, in %most real cases
Since in many real-world % application 
scenarios, such as network dismantling~\cite{holme2002attack}, it is the relative importance of nodes (as measured by BC), and moreover, the top-$N\%$ nodes, that %matters
serve as the key to % resolve 
solving the problem, rather %not
than the exact BC values \cite{kourtellis2013identifying,mahyar2018compressive,kumbhare2014efficient}. 
%\muhao{This needs citations.}. \yuhui{Second says nothing} 
%Second,in many applications, identifying the top N\% most important nodes is more practical than precisely ordering the nodes based on their exact betweenness centrality \muhao{This also needs citations.}.
%\muhao{(citation is needed for this claim)}. 
%The most
A straightforward way to obtain the top-$N\%$ highest BC nodes %would be
is to compute the BC values of all nodes using exact or approximation algorithms, and then identify top-$N\%$ among them. %and 
%then sort them. 
%identifying the top N\% based on the BC values of all nodes given by an exact or approximation BC computing algorithms. 
%\muhao{I rephrased the sentence to this, becasue even though the BC nodes are computed, the optimal solution is not to sort it, but is to use a N-sized min-heap to do a one-path scan.}
However, the time complexity of these solutions is unaffordable for large networks %systems 
with millions of nodes.

% To mitigate these issues, recently, extensive efforts have been made on approximation algorithms for BC~\cite{brandes2007centrality,bader2007approximating,pfeffer2012k,everett2005ego,bergamini2016approximating,riondato2016fast,riondato2018abra,erdHos2015divide,sariyuce2013shattering,geisberger2008better,khopkar2016penalty,mahmoody2016scalable,yoshida2014almost}. 
% Most existing approaches are sampling-based, which estimate BC for each node according to a set of sampled nodes or subgraphs. These approaches often tackle the trade-off in balancing computational efficiency and predictive accuracy. However, most existing approaches are still inefficient to handle large-scale networks, let alone maintaining updates of BC on dynamic networks. 

%solution overview
To address this issue, we propose to %identify high BC nodes by turning it into a learning problem.
transform the problem of identifying high BC nodes into a learning problem.
The goal is to learn an \emph{inductive} BC-oriented operator 
%\yuhui{BC operator means BC is what to be operated, it's misleading. It's not appropriate to add "BC" in front of any term, the same as BC vector} 
that is able to directly map any network topology into a %BC vector
ranking score vector, where each entry denotes the %BC ranking score
ranking score of a node with regard to its BC. %for the corresponding node, and note 
Note that this score does not need to approximate the exact BC value. Instead, it indicates the relative order of nodes with regard to BC. We employ an encoder-decoder framework where the encoder maps each node to an embedding vector  %capturing
which captures the essential structural information related to BC computation, and the decoder maps embedding vectors into BC ranking scores.

%\yuhui{Use Challenge is better} 
One major challenge here is: \emph{How can we represent the nodes and the network?} Different network embedding approaches have been proposed to map the network structure into a low-dimensional space where the proximity of nodes or edges are captured~\cite{hamilton2017representation}. These embeddings have been exploited as features for various downstream prediction tasks, such as multi-label node classification~\cite{perozzi2014deepwalk,chen2018enhanced, fan2014fuzzy}, link prediction~\cite{grover2016node2vec,fan2017efficient} and graph edit distance computation~\cite{bai2018graph}. Since BC is a measure highly related to the network structure, we propose to design a network embedding model that can be used to predict BC ranking scores. To the best of our knowledge, this work represents the first effort for this topic.

We propose \textbf{\model}\ %, abbreviated from Deep BC Ranker,
(\ul{\textbf{D}}eep \ul{\textbf{r}}anker for \ul{\textbf{BC}}), a graph neural network-based ranking model to identify the high BC nodes. At the encoding stage, \model~captures the structural information for each node in the embedding space. At the decoding stage, it leverages the embeddings to compute BC ranking scores which are later used to find high BC nodes. More specifically, the encoder part is designed in a neighborhood-aggregation fashion, and the decoder part is designed as a multi-layer perceptron (MLP). The %model's
model parameters are trained in an end-to-end manner, where the training data %are
consist of different synthetic graphs %with their nodes' ground truth BC values
labeled with ground truth BC values of all nodes. The learned ranking model can then be applied to any unseen networks. \nop{Note that, our model is inductive, namely, the learned BC-oriented operator can be applied to any unseen network.} %\YS{Wait to see why the proposed encoder is good to capture structure information related to BC.}

%Empirically,
We conduct extensive experiments on synthetic networks of %seven different sizes
a wide range of sizes and five large real networks from different domains. The results show \model~can %generate embeddings that capture %BC proximity between nodes,
effectively induce the
partial-order relations of nodes regarding their BC from the embedding space.
Our method achieves at least comparable accuracy on both synthetic and real-world networks to state-of-the-art sampling-based baselines, and much better performance than the top-$N\%$ dedicated baselines~\cite{borassi2016kadabra} and traditional node embedding models, such as Node2Vec~\cite{grover2016node2vec}. For the running time, our model is far more efficient than sampling-based baselines, node embedding based regressors, and is comparable to the top-$N\%$ dedicated baselines. 
%\YS{better name for the top-k baseline}
%\YS{XX times speed up to the second best?}\FC{OK, I will later add them}
%This could relieve the recursive neighborhood expansion challenge incurred by training large networks\cite{chen2018fastgcn}, and speed up the inference time on large networks, since no training process is needed on them.

%To summarize, we make the following contributions:
The main contributions of this paper are summarized as follows:
\begin{enumerate}
    \item We transform the problem of identifying high BC nodes into a learning problem, where an inductive embedding model is learned to capture the structural information of unseen networks to provide node-level features that help BC ranking. %where a BC-oriented operator that maps a network into a BC vector is learned and can be applied to unseen networks.
    %and resolves it with an inductive learning method that excels in both accuracy and computational efficiency.
    \item We propose a graph neural network based encoder-decoder model, \model, to rank nodes specified by their BC values. The model first encodes nodes into embedding vectors and then decodes them into BC ranking scores, which are utilized to identify the high BC nodes.
    \item We perform extensive experiments on both synthetic and real-world datasets in different scales and in different domains. Our results demonstrate that \model~performs on par with or better than state-of-the-art baselines, while reducing the execution time significantly.
    %\item %\nop{We test our approach in the network dismantling task, where we repeatedly remove the node with the highest approximate BC score until the network is completely broken down. Extensive evaluations on nine p2p networks demonstrate that \model\ based attack strategy outperforms all baselines, in terms of both robustness scores and running time.} 
    % We have demonstrated the generalizability of our model by showing its capability of transferring knowledge across different types of networks, which makes it practical to train the model on small-scale synthetic networks and apply it to large-scale real-world networks. 
    %\YS{revision needed, as we still need to make sure the training networks have to be PL-cluster style networks.}
\end{enumerate}

The % remainder part 
rest of the paper is organized as follows. We systematically review related work in Section 2. %including BC approximation algorithms, network embedding models and learning-to-rank approaches. 
After that, we introduce the architecture of \model~ in detail in Section 3. %Then we test our model on both synthetic networks and large real-world networks, and analyze the results in Section 4.
Section 4 presents the evaluation of our model on both synthetic networks and large real-world networks. We discuss some observations which intuitively explain why our model works well %on this problem 
in section 5. Finally, we conclude the paper in Section 6. 
%To demonstrate the adaptiveness and  usefulness of \model, we also apply \model\space to the task of network dismantling, and evaluate its performance in Section 5\FC{Later check it}. 

\section{Related Work}
\subsection{Computing Betweenness Centrality}
Since the time complexity of the exact BC algorithm, Brandes Algorithm~\cite{brandes2001faster}, is prohibitive for large real-world networks, many approximation algorithms which trade accuracy for speed have been developed. A general idea of approximation is to use a \emph{subset} of pair dependencies instead of the complete set required by the exact computation. \citeauthor{riondato2016fast}~\cite{riondato2016fast} introduce the Vapnik-Chervonenkis (VC) dimension to compute the sample size that is sufficient to obtain guaranteed approximations of all nodes' BC values. To make the approximations correct up to an additive error $\lambda$ with probability $\delta$, the number of samples is $\frac{c}{\lambda^2}(\lfloor log(VD-2) \rfloor + 1 + log\frac{1}{\delta})$, where $VD$ denotes the maximum number of nodes on any shortest path.
\citeauthor{riondato2018abra}~\cite{riondato2018abra} use adaptive sampling to obtain the same probabilistic guarantee as \cite{riondato2016fast} with often smaller sample sizes. \citeauthor{borassi2016kadabra}~\cite{borassi2016kadabra} follow the idea of adaptive sampling and propose a balanced bidirectional BFS, reducing the time for each sample from $\Theta(|E|)$ to $|E|^{\frac{1}{2}+O(1)}$.

To identify the top-$N\%$ highest BC nodes, both \cite{riondato2016fast} and \cite{riondato2018abra} need another run of the original algorithm, which is costly on real-world large networks.
\citeauthor{kourtellis2013identifying}~\cite{kourtellis2013identifying} introduce a new metric and show empirically that nodes with high this metric have high BC values, and then focus on computing this alternative metric efficiently. \citeauthor{chung2014finding}~\cite{chung2014finding} utilizes the novel properties of bi-connected components to compute BC values of a part of vertices, and employ an idea
of the upper-bounding to compute the relative partial order of the vertices regarding their BCs. \citeauthor{borassi2016kadabra}~\cite{borassi2016kadabra} propose a variant for efficiently computing top-$N\%$ nodes, which allows bigger confidence intervals for nodes whose BC values are well separated. However, as we show in our experiments, these methods still cannot achieve a satisfactory trade-off between accuracy and efficiency, which limits their use in practice.

\subsection{Network Embedding}
Network embedding has recently been studied to characterize a network structure to a low-dimensional space, and use these learned low-dimensional vectors for various downstream graph mining tasks, such as node classification and link prediction~\cite{grover2016node2vec}.
%, and anomaly detection \cite{hu2016embedding}. 
%such that structural information such as high-order proximity of nodes is preserved in the embedding space. Once achieving that, these learned low-dimensional vectors can be used as input features in various machine learning models to handle different graph mining tasks, such as node classification \cite{perozzi2014deepwalk}, link prediction \cite{liben2007link},and anomaly detection \cite{hu2016embedding}. This differs from conventional paradigms which strongly rely on extracting hand-crafted features based on network properties~\cite{henderson2012rolx}. Network embedding models automate the process by casting feature extraction as a feature learning problem~\cite{grover2016node2vec}, and they have been demonstrated to be effective in various network analysis tasks~\cite{perozzi2014deepwalk, grover2016node2vec, tang2015line}. \muhao{The last sentence looks redundant to the previous two.}
Current embedding-based models share a similar encoder-decoder framework~\cite{hamilton2017representation}, where the encoder part maps nodes to low-dimensional vectors, and the decoder %decodes
infers network structural information from the encoded vectors. Under this framework, there are two main %stream
categories of approaches. %, first
The first category is the direct encoding approach, where the encoder function is just an embedding lookup function, and %nodes' embedding 
the nodes are parameterized as embedding vectors to be optimized directly. The decoder function is typically based on the inner product of embeddings, which seeks to obtain deterministic measures such as network proximity~\cite{tang2015line} or statistics derived from random walks~\cite{grover2016node2vec}. This type of models suffers from several major limitations. First, it does not consider the node attributes, which are quite informative in practice. %Second, it is not permutation invariant. In other words, if we change the order of the nodes, the output of the model will be completely different. 
Second, no parameters are shared across nodes, and the number of parameters necessarily grows as $O(|V|)$, which is computationally inefficient. Third, it is not able to handle previously unseen nodes, which prevents their application on dynamic networks.

\nop{These direct %encoded
encoding models suffer from high computational cost, since the number of parameters to be optimized equals to $|V|d$, where $d$ is the dimension of the embedding vector and $V$ is the total number of nodes. Besides, due to that the %node's
nodes' embeddings are directly optimized, %it's
it is hard to integrate %nodes' attributes
the attributes of nodes in the learning process, which are often highly informative with respect to the %nodes' 
positions and roles of nodes in the network. Furthermore, direct encoding methods %fail to generate
fall short of %generating embeddings for previously
embedding unseen nodes, which prevents their application in evolving networks.
}

To address the above issues, the second category of models have been proposed, which %we called
are known as the neighborhood aggregation models \cite{kipf2016semi,hamilton2017inductive}. %here. 
For these models, the encoder function is to iteratively aggregate %neighborhood's 
embedding vectors from the neighborhood, which are initialized as node feature vectors, and followed by a non-linear transformation operator. When nodes are not associated with any attributes, simple network statistics based features are often adopted, such as node degrees and local clustering coefficients. The decoder function can %be either
either be the same as the ones in the previous category, or be integrated with task-specific supervisions~\cite{chen2017task}. This type of embedding framework turns out to be more effective in practice, due to its flexibility to incorporate node attributes and apply deep, nonlinear transformations, and %integrating with
adaptability to downstream tasks. In this paper, we follow the neighbor aggregation model for BC approximation.  

\nop{based on repeating several iterations of neighborhood aggregation operations with node attributes as input features. For those networks without attributes, simple network statistics are often adopted, such as degrees. The decoder function can either be the same as the previous decoders or be incorporated with task-specific supervisions. In this paper, we utilize the neighborhood aggregation models for BC approximation.  
}
%A typical neighborhood aggregation %scheme
%function can be %written
%defined as follows. Let $h_v^{(l)}$ denote the embedding vector for node $v$ at layer $l$, it will be defined via two steps: 

Let $h_v^{(l)}$ denote the embedding vector for node $v$ at layer $l$, A typical neighborhood aggregation function can be defined in two steps: (a) the aggregation step that aggregates %$v$'s neighbors' embedding vectors
the embedding vectors from the neighbors of $v$ at layer $l-1$ and (b) the transformation step that combines %transforms %$v$'s last layer embedding
the embedding of $v$ in the last layer and %its neighbors' aggregated embedding
the aggregated neighbor embedding %into
to %its current embedding:
the embedding of $v$ of the current layer.
\nop{For a $K-$layer model, the $l-$th layer ($l=1, ..., K$) updates $h_v^{(l)}$ for $v\in V$ simultaneously as:}
\begin{align}
    & h_{N(v)}^{(l)} = AGGREGATE(\{h_u^{(l-1)}, \forall u \in N(v)\}) \\
    & h_v^{(l)} = \sigma(W_l \cdot COMBINE(h_v^{(l-1)}, h_{N(v)}^{(l)})) \label{NG-genernal}
\end{align}
where $N(\cdot)$ denotes a set of neighboring nodes of a given node,  $W_l$ is a trainable weight matrix of the $l-$th layer shared by all nodes, and $\sigma$ is %a non-linear
an activation function, e.g. ReLU. $AGGREGATE$ is a function that aggregates information from local neighbors, while $COMBINE$ is a function that combines %the node $v$'s representation
the representation of node $v$ at layer $l-1$, i.e. $h_v^{(l-1)}$, with the aggregated neighborhood representation at layer $l$, $h_{N(v)}^{(l)}$. The $AGGREGATE$ function and the $COMBINE$ function are defined specifically in different models.

Based on the above neighborhood aggregation framework, many models focus on addressing the following four questions:
\begin{itemize}[leftmargin=1em]
    \item \textbf{How to define the neighborhood?} Some directly use all adjacent nodes as neighbors~\cite{kipf2016semi}, while others just sample some of them as neighbors~\cite{hamilton2017inductive}.
    \item \textbf{How to choose the AGGREGATE function?} Based on the definition of neighbors, there are multiple aggregation functions, such as sum~\cite{kipf2016semi}, mean~\cite{hamilton2017inductive}. \citeauthor{hamilton2017inductive} \cite{hamilton2017inductive} introduce some %other 
    pooling functions, like LSTM-pooling and max-pooling. ~\citeauthor{velickovic2017graph}~\cite{velickovic2017graph} propose an attention-based model to learn weights for neighbors and use the weighted summation as the aggregation.
    \item \textbf{How to design the COMBINE function?} Summation \cite{kipf2016semi} and concatenation~\cite{hamilton2017inductive} are two typical functions, and~\citeauthor{li2015gated}~\cite{li2015gated} use the Gated Recurrent Units (GRU).
    \item \textbf{How to handle different layers' representations?} Essentially, deeper models get access to more information. However, in practice, more layers, even with residual connections, do not perform better than less layers (2 layers)~\cite{kipf2016semi}.~\citeauthor{xu2018representation}~\cite{xu2018representation} point out that this may be that real-world complex networks possess locally varying structures. They propose three layer-aggregation mechanisms: concatenation, max-pooling and LSTM-attention, to enable adaptive, structure-aware representations.
\end{itemize}
\noindent Nevertheless, these models cannot be directly applied to our problem. Hence, we carefully design a new embedding function such that the embeddings can preserve the essential information related to BC computation.

% \subsection{Learning-to-Rank}
% Learning to rank is to construct a ranking model %for ranking in search, 
% in a data-driven way. There are many algorithms proposed for this field. Current models can be categorized as point-wise approaches, pairwise approaches and list-wise approaches, based on different loss functions in learning~\cite{liu2009learning}. The pairwise and listwise models are usually better than the pointwise algorithms, since the key issue of ranking is to determine the relative orders between items, rather not the relevance between items. The former is just the goal of pairwise and list-wise algorithms, while the point-wise models optimize the latter. The representative pairwise models include RankNet~\cite{burges2005learning}, LambdaMART~\cite{burges2010ranknet}. Here we utilize the pairwise loss in our model, since it can capture more global ranking information compared to point-wise loss, and make the training more stable and converge faster compared to list-wise loss.
%\vspace{-1em}
%\section{Our Method}
\section{Proposed Method: \model}
In this section, we %explain our method,
introduce the proposed model, namely \model, for BC ranking. %Firstly, we define some notations, and then we explain the framework of \model and its training algorithms.
We begin with the preliminaries and notations. 
Then we introduce the architecture of \model \space in detail, as well as its training procedure. Finally, we analyze the time complexity for training and inference.

\subsection{Preliminaries}
Betweenness centrality (BC) %is essential centrality measure, which 
indicates the importance of individual nodes based on the fraction of shortest paths that pass through them. %In this paper, we focus on the node BC. 
Formally, the normalized BC value $b(w)$ of a node $w$ is defined:
\begin{equation}\label{definition}
    \begin{aligned}
        b(w) = \frac{1}{|V|(|V|-1)}\sum_{u \ne w \ne  v}\frac{\sigma_{uv}(w)}{\sigma_{uv}}
    \end{aligned}
\end{equation}
where $|V|$ denotes the number of nodes in the network, $\sigma_{uv}$ denotes the number of shortest paths from $u$ to $v$, and $\sigma_{uv}(w)$ denotes
the number of shortest paths from $u$ to $v$ that pass through $w$.%\YS{is $u\ne w \ne v$ a standard way to denote the three nodes are not identical to each other? }

\begin{figure}[t!]
\centering
\includegraphics[width=0.7\linewidth]{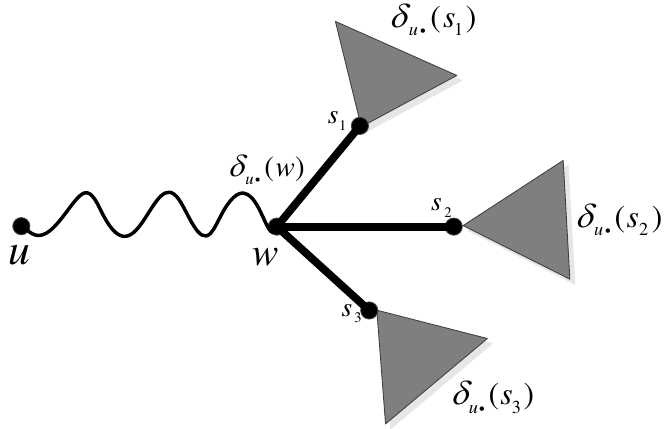}
\caption{Illustration of Eq. (\ref{update}). $w$ lies on shortest paths from source $u$ to $s_1$, $s_2$ and $s_3$, which are %$w$'s
the successors of $w$ in the tree rooted at $u$. The figure is adapted from Figure 1 in~\cite{brandes2001faster}.}\label{brandes_illus}
\end{figure}

The Brandes Algorithm \cite{brandes2001faster} is the asymptotically fastest algorithm for computing the exact %betweenness centrality
BC values of all nodes in a network. %For any %triplet of
Given three nodes $u$, $v$ and $w$, pair dependency, denoted by $\delta_{uv}(w)$, and source dependency, denoted by $\delta_{u \centerdot }(w)$, are defined as:
\begin{equation}
    \begin{aligned}
        \delta_{uv}(w)=\frac{\sigma_{uv}(w)}{\sigma_{uv}} \quad and\quad \delta_{u\centerdot}(w)=\sum_{v \ne w}{\delta_{uv}(w)}
    \end{aligned}
\end{equation}
With the above notations, Eq. (\ref{definition}) can be equivalently written as:
\begin{equation}
    \begin{aligned}
        b(w)=\frac{1}{|V|(|V| - 1)}\sum_{u\ne w}{\sum_{v\ne w}{\delta_{uv}(w)}}=\frac{1}{|V|(|V|-1)}\sum_{u\ne w}\delta_{u\centerdot}(w)
    \end{aligned}
\end{equation}
%Brandes
\citeauthor{brandes2001faster} proves that $\delta_{u\centerdot}(w)$ can be computed as follows:
\begin{equation}\label{update}
    \begin{aligned}
        \delta_{u\centerdot}(w)=\sum_{s: w \in P_u(s)}\frac{\sigma_{uw}}{\sigma_{us}} \cdot (1 + \delta_{u\centerdot}(s))
    \end{aligned}
\end{equation}
where $P_u(s)$ denotes the predecessors of $s$ in the shortest-path tree rooted at $u$. Eq. (\ref{update}) can be illustrated in Figure \ref{brandes_illus}. 
The algorithm performs a two-phase process. The first phase executes a shortest-path algorithm from $u$ to compute $\sigma_{uv}$ and $P_u(v)$ for all nodes $v$ with $v\ne u$. The second phase performs reversely from the leaves to the root and uses Eq. (\ref{update}) to compute $\delta_{u\centerdot}(w)$. The whole process
runs in \nop{$O(mn)$ and $O(mn+n^2logn)$ time for unweighted and weighted networks, respectively.}$O(|V||E|)$ for unweighted networks and $O(|V||E|+|V|^2log|V|)$ for weighted networks, respectively, where $|V|$ and $|E|$ denote the number of nodes and edges in the network.

\subsection{Notations}
%Let $G(V,E)$ be a %single 
Let $G=(V, E)$ be a
network with node features (attributes, structural features or arbitrary constant vectors) $X_v \in \mathbb{R}^{c}$ for $v \in V$, where $c$ denotes the dimension of input feature vectors, and $|V|$, $|E|$ be the number of nodes and edges respectively.
%Notation 
$h_v^{(l)} \in \mathbb{R}^{p} \ (l=1, 2, ..., L)$ denotes the embedding of node $v$ at the $l-$th layer of the model, where $p$ is the dimension of hidden embeddings, which we assume to be the same across different layers for the sake of simplicity. We use $[d_v,1,1]$ as node $v$'s initial feature $X_v$, and let $h_v^{(0)} = X_v$. The neighborhood of node $v$,  $N(v)$, %is
is defined as all the nodes that %have an edge to
are adjacent to $v$, and $h_{N(v)}^{(l)}$ denotes the aggregated neighborhood representation output by the $l-$th layer of the model. %\YS{(1) BC has nothing to do with node features. It might cause confusion when you say it is a graph with node features. You probably want to mention it later for the initialization. (2) It turns out that if starting with constant initialization, the embedding for nodes with structural equivalence will be the same, which is good for BC computation. If we use other derived network features, we need to show that it will enhance this phenomena.}\FC{Fine, I will consider it later}

\subsection{\model}
\begin{figure}[t!]
\centering
\includegraphics[width=\linewidth]{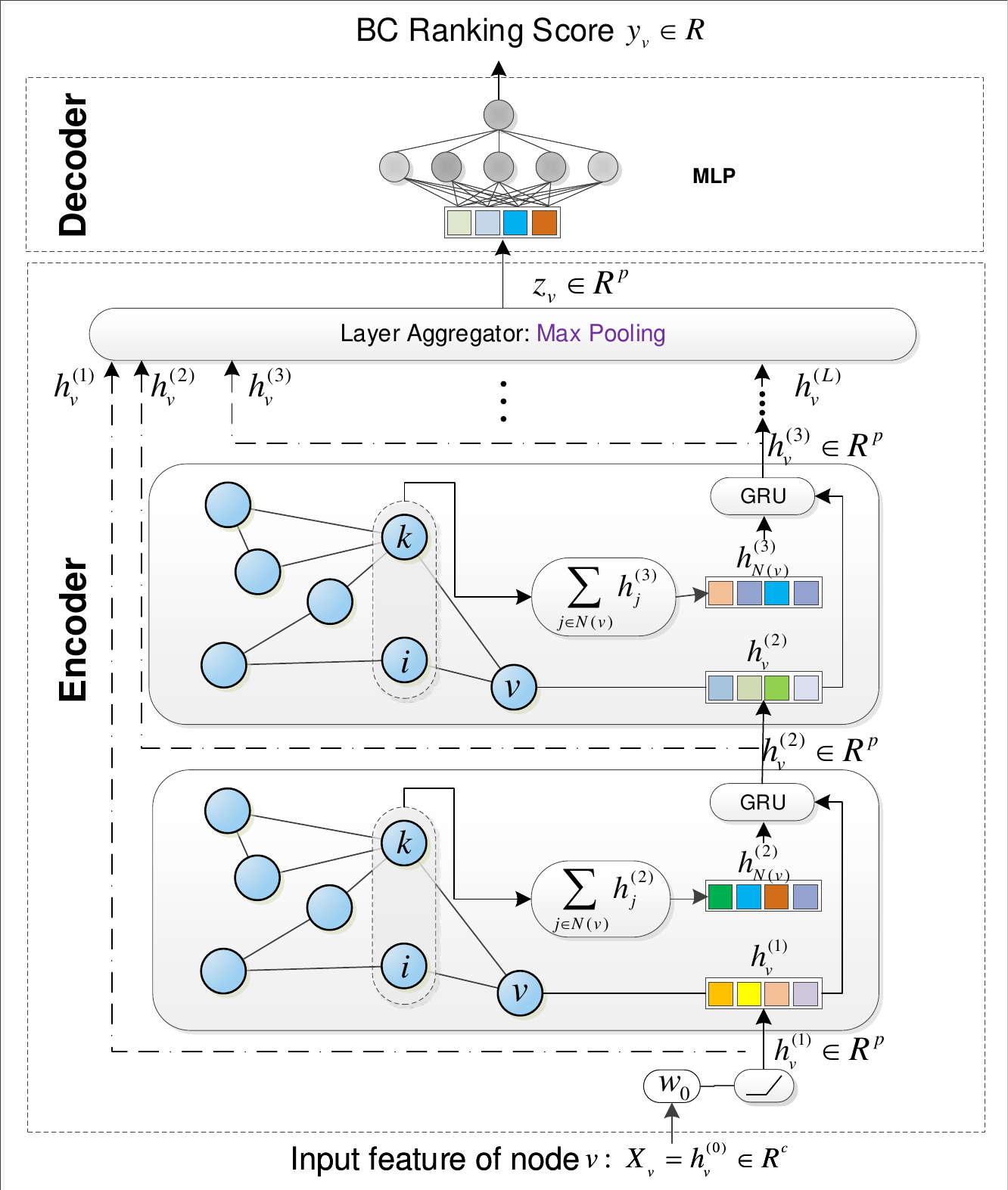}
\caption{The encoder-decoder framework of \model}\label{architecture}.%\YS{Plot needs to be more professional. Figure 2 is a good one.}}
\vspace{-1em}
\end{figure}

We now introduce our model \model\ in detail, which %also 
follows an encoder-decoder framework. Figure \ref{architecture} illustrates the main architecture of \model. It consists of two %main parts:
components: 1) %Encoder part,
the \emph{encoder} that %to 
generates node embeddings, and 2) %Decoder part, to
the \emph{decoder} that maps the embeddings to scalars for ranking nodes specified by BC values.
%to %\yuhui{not BC values} scalars to approximate %nodes' 
% BC values.

\subsubsection{\textbf{The Encoder.}}
%Different from existing embedding methods, which seek to encode nodes into embedding vectors that can preserve the topological proximity between nodes, our embedding is able to preserve \yuhui{?} the similarity in terms of BC score.
As is indicated in %(\ref{brandes})
Eq.~(\ref{update}) and Figure~\ref{brandes_illus}, computing a node's exact BC value needs to iteratively aggregate its neighbors information, which is similar as the neighbor aggregation schema in graph neural networks. Therefore, we explore to choose GNNs as the encoder, and its inductive settings enable us to train on small-scale graphs and test directly on very large networks, which is the main reason for the efficiency of \model. For the specific design, we consider the following four components:
 %$\delta_{u\centerdot}(w) = \sum_{s:w\in P_u (s)} \frac{\sigma_{uw}}{\sigma_{us}}(1+\delta_{u\centerdot}(s))$, where $w$ is one of the predecessors of node $s$ on shortest paths rooted at $u$, thus $s$ is adjacent to $w$, $\delta_{u\centerdot}(w)$ and $\delta_{u\centerdot}(s)$ are related. This kind of relation is shown in Figure \ref{brandes_illus}, which further indicates that there exists %the neighborhood aggregation scheme in the process of computing exact BC, a correlation between the BC value of node $w$ and the BC values of $w$'s neighbors. %similar neighbor aggregations as the neighborhood aggregation-based embedding model. %Another reason is that this model Such observation inspires a neighborhood-aggregation encoder, and note that this encoder can naturally be employed in inductive settings, where we %train the model on small networks and test on large ones, since %it's
%it is prohibitive to compute
%the exact BC in large networks, which results in the lack of ground truth for training on these networks. directly apply the trained model to previously unseen real-world large networks.\nop{ and get the approximate BC values much faster than any approximation algorithm.}
%We design the neighborhood-aggregation encoder by defining the following four components: %questions of neighborhood aggregation based embedding.

\noindent \textbf{Neighborhood Definition.}
%In the form of (\ref{brandes}),
According to Eq.~(\ref{update}), the exact computation of the %node's 
BC value of a node relies on its immediate neighbors. Therefore, we %will 
include all immediate neighbors of a node for full exploitation. Note that our model will be trained only on small networks, and utilizing all adjacent neighbors would not incur expensive computational burdens\nop{in mini-batch training due to the recursive expansion of neighborhoods across layers}.

\noindent \textbf{Neighborhood Aggregation.}
%Still refer to (\ref{brandes}),
In Eq.~(\ref{update}), each neighbor is weighted by the term $\frac{\sigma_{uw}}{\sigma_{us}}$, where $\sigma_{ij}$ denotes the number of shortest paths from $i$ to $j$. As the number of shortest paths between two nodes is expensive to compute in large networks, we propose to use a weighted sum aggregator to aggregate neighbors, \nop{which serves the purpose of the weighted sum function} which is defined as follows \nop{\YS{previously you use``equivalent''. I did not see why they are equivalent.}}:
\begin{equation}
\begin{aligned}
    h_{N(v)}^{(l)} = \sum_{j\in N(v)} \frac{1}{\sqrt{d_v+1} \cdot \sqrt{d_j+1}} h_{j}^{(l-1)} \label{brandes}
\end{aligned}
\end{equation}
where $d_v$ and $d_j$ denote the degrees of node $v$ and node $j$, $h_{j}^{(l-1)}$ denotes node $j$'s embedding output by the $(l-1)$-th layer of the model.
%and $h_{j}^{(0)}=X_j$ is
%the input feature of node $j$. 
%\YS{The equation implies higher degree node is less important. Not sure why it is the case for BC computation. One may have opposite intuition. }
%For networks without nodes' input features, we can use one-hot indicators or constant vectors instead. 
For the aggregator, the weight here is determined by the node's degree, which is both efficient to compute and effective to describe its topological role in the network. 
%\yuhui{Why topological role indicates weight? more intuition.}%And attention mechanisms\cite{velickovic2017graph} can also be used to learn weights automatically, we put it in the future work.
Note that the attention mechanism~\cite{velickovic2017graph} can also be used to automatically learn neighbors weights, which we leave as future work.

\noindent \textbf{COMBINE Function.} 
% Each node is influenced by its previous state and the state of its neighboring nodes in order to propagate information to the whole structure. The $COMBINE$ function just deals with the combination of these two states. 
The $COMBINE$ function deals with the combination of the neighborhood embedding generated by the current layer, and the embedding of the node itself generated by the previous layer. Most existing models explore the forms of sum~\cite{kipf2016semi} and concatenation~\cite{hamilton2017inductive}. %function, functions, which can be regarded as simple weighted average or `skip connection'.
%~\YS{I do not understand why it is called skip connection. If it is not essential to explain the whole idea, remove it; otherwise, explain it.}\muhao{Do you mean ``residual shortcuts''?} between the adjacent two layers. \yuhui{In this way it seems to me combine function is a little similar to layer-wise aggregation. I think you should demonstrate the difference better in the layer aggregation part}
Here we propose to use the GRU. Let the neighborhood embedding $h_{N(v)}^{(l)}$ generated by the $l$-th layer be the input state, and node $v$'s embedding $h_{v}^{(l-1)}$ generated by the $(l-1)$-th layer be the hidden state, then the embedding of node $v$ at the $l$-th layer can be written as $h_{v}^{(l)} = GRUCell(h_{v}^{(l-1)}, h_{N(v)}^{(l)})$. The GRU utilizes the gating mechanism. The update gate $u_l$ helps the model determine how much past information needs to be passed along to the future, and the reset gate $r_l$ determines how much past information to be forgotten. Specifically, the GRU transition $GRUCell$ is given as the following:
\begin{align}
& u_l = sigmoid(W_1 h_{N(v)}^{(l)} + U_1 h_{v}^{(l-1)}) \label{update_gate} \\
& r_l = sigmoid(W_2 h_{N(v)}^{(l)} + U_2 h_{v}^{(l-1)}) \label{reset_gate}  \\
& f_l = tanh(W_3 h_{N(v)}^{(l)} + U_3 (r_l \odot h_{v}^{(l-1)})) \\
& h_{v}^{(l)} = u_l \odot f_l + (1-u_l) \odot h_{v}^{(l-1)}
\end{align}
%Here
\noindent where $\odot$ represents the element-wise product. 
%\YS{$\sigma$ has been used to denote both number of shortest paths and sigmoid function.} 
%We initially experimented with a vanilla recurrent neural network-style update, but in preliminary experiments we found this GRU-like propagation step to be more effective.
%\YS{what is the intuition here to use reset gate and update gate? Whay turn this problem into a sequence problem?}\yuhui{Probably you can refer to the original paper who uses GRU}
With GRU, our model can learn to decide how much proportion of the features of distant neighbors should be incorporated into the local feature of each node. In comparison with other $COMBINE$ functions, GRU offers a more flexible way for feature selection, and gains a more accurate BC ranking in our experiments. 
%\YS{We need to show why the combination can be considered as a sequence encoding problem.}

\noindent \textbf{Layer Aggregation.}
%Conventional neighborhood aggregation models repeat the same number of iterations (layers) for all nodes, which may lead to {\color{red} influence distributions} of %verydrastically different locality~\cite{xu2018representation}, \yuhui{It would be strange if you just cited these terms without their original context. I think it would be OK to omit these terms and directly begin with examples. Some nodes with high BC balabalabala...}\YS{agree with Yuhui.} 
As~\citeauthor{li2018deeper}~\cite{li2018deeper} point out that each propagation layer is simply a special form of smoothing which mixes the features of a node and its nearby neighbors. However, repeating the same number of iterations (layers) for all nodes would lead to over-smoothing or under-smoothing for different nodes with varying local structures. 
%since many real-world networks possess strongly varying local structures.
Some nodes with high BC values are located in the core part or within a few hops away from the core, and their neighbors thus can expand to nearly the entire network within few propagation steps. However, for those with low BC values, like nodes with degree one, their neighborhoods often cover much %less
fewer nodes within the same %iterations
steps. This implies that the same number of iterations for all nodes may not be reasonable, %
and wide-range or small-range propagation combinations based on %nodes' 
local structures may be more desirable. %As such, 
Hence, \citeauthor{xu2018representation}~\cite{xu2018representation} explore three layer-aggregation approaches: concatenation, max-pooling and LSTM-attention respectively. In this paper, we %exploit
investigate the max-pooling aggregator, which is defined as an element-wise
operation
$max(h_{v}^{(1)}, ..., h_{v}^{(L)})$ %, which
to select the most informative (using $max$ operator) layer for each feature coordinate. %It is
This mechanism is adaptive and easy to be implemented, and it introduces no additional parameters to %learn.
the learning process. In our experiments, we find it more effective than the other two layer aggregators for BC ranking. Practically, we set the maximum number of layers to be 5 when training, and more layers lead to no improvement.
%\YS{How many layers are used in practice? What would be its impact to the performance?}

%Put the above four components together,
Combining the above four components, we %can get
have the following encoder function: 
\begin{equation}
    z_v = ENC(A, X; \Theta_{ENC})
\end{equation}
where $A$ denotes the adjacency matrix, $X$ denotes node features, and $\Theta_{ENC}=\{W_0\in \mathbb{R}^{c \times p}, W_1, U_1, W_2, U_2, W_3, U_3 \in \mathbb{R}^{p \times p}\}$. %details are illustrated in Algorithm \ref{graphEmbed}.
The detailed encoding process is shown in Algorithm~\ref{graphEmbed}.

\begin{algorithm}[!htb]
\footnotesize\caption{\model \space encoder function}
\begin{algorithmic}[1]\label{graphEmbed}
\REQUIRE
Network $G=(V,E)$; input features $\{X_{v}\in\mathbb{R}^{c}, \forall v \in V\}$; depth $L$; weight matrices $W_{0} , W_1, U_1, W_2, U_2, W_3, U_3$.
\ENSURE %output
Vector representations $z_{v}$, $\forall v \in V$.
\STATE Initialize $h^{(0)}_{v} = X_v$;
\STATE $h^{(1)}_{v} = ReLU(W_0 h^{(0)}_v)$, $h_{v}^{(1)} = h_{v}^{(1)} / \|h_{v}^{(1)}\|_{2}, \forall v \in V$;
\FOR{$l=2$ to $L$}
\FOR{$v \in V$}
\STATE $h_{N(v)}^{(l)} = \sum_{j \in N(v)} \frac{1}{\sqrt{d_v+1}\cdot\sqrt{d_j+1}} h_{j}^{(l-1)}$;
\STATE $h_{v}^{(l)} = GRUCell(h_{v}^{(l-1)}, h_{N(v)}^{(l)})$;
\ENDFOR
\STATE $h_{v}^{(l)} = h_{v}^{(l)} / \|h_{v}^{(l)}\|_{2}, \forall v \in V$;
\ENDFOR
\STATE $z_{v} = max(h_{v}^{(1)},h_{v}^{(2)},...,h_{v}^{(L)} ), \forall v \in V$;
\end{algorithmic}
\end{algorithm}

\subsubsection{\textbf{The Decoder.}} The decoder is implemented with a two-layered MLP which maps the embedding $z_v$ to the approximate BC ranking score $y_v$:
\begin{equation} \label{decoder}
    y_v = DEC(z_v; \Theta_{DEC}) = W_5 ReLU(W_4 z_v)
\end{equation}
\noindent where $\Theta_{DEC}=\{W_4\in \mathbb{R}^{p\times q}, W_5\in \mathbb{R}^{q}\}$, $p$ denotes the dimension of $z_v$, and $q$ is the number of hidden neurons.

\subsection{Training Algorithm}
There are two sets of parameters to be learned, including $\Theta_{ENC}$ and $\Theta_{DEC}$. We use the following pairwise ranking loss to update these parameters.

Given a node pair $(i,j)$, suppose the ground truth BC values are $b_i$ and $b_j$, respectively, %our model's predicted BC ranking scores are $y_i$ and $y_j$
our model predicts two corresponding BC ranking scores $y_i$ and $y_j$. %Define
Given $b_{ij}\equiv b_i-b_j$, %and $y_{i,j}\equiv y_i-y_j$, since we seek to preserve the relative rank order specified by the ground truth, the cross-entropy cost function is utilized for it can be cast as a binary classification problem:
since we seek to preserve the relative rank order specified by the ground truth,
our model learn to infer $y_{ij}\equiv y_i-y_j$ based on the following binary cross-entropy cost function,
\begin{equation}
    C_{i,j} = -g(b_{ij})*log\sigma(y_{ij}) - (1-g(b_{ij}))*log(1-\sigma(y_{ij}))
\end{equation}
where $g(x)=1/(1+e^{-x})$. As such, the training loss is defined as:
\begin{equation}
    Loss = \sum_{i,j\in V}C_{i,j} \label{loss}
\end{equation}
% Note that node $i$ and $j$ are randomly sampled from the whole graph nodes. On average, each node is sampled five times. 
In our experiments, we randomly sample $5|V|$ source nodes and $5|V|$ target nodes with replacement, forming $5|V|$ random node pairs to compute the loss.
%More sampling details see in Appendix \ref{appendix_nodesample}. 
Algorithm \ref{train_algo} describes the training algorithm for \model.

% Since we train \model~on small networks (in our experiments, size ), it is affordable to calculate the exact BC values of all nodes with Brandes algorithm~\cite{brandes2001faster}. With the exact values as the ground truth, these parameters can be trained easily by minimizing the mean squared error between the predicted BC values and the ground truth:
% \begin{equation}
%     Loss = \sum_{v\in V} (y_v - b_v)^2 \label{loss}
% \end{equation}
% where $y_v$ is the predicted BC value for node $v$, %$b_v$ is the exact BC as the ground truth.
% and $b_v$ is the exact BC value. \YS{new loss.}

%We apply several transformations on the training labels based on some characteristics of BC values to improve the accuracy. First, due to the empirical power-law distribution nature~\cite{goh2001universal} in most real-world datasets, we smooth the BC values using logarithmic scaling to make them easier to be learned. Second, there are many nodes with BC values of zero in the network, such as nodes with degree of one and nodes with local clustering coefficient of one. These nodes are easy to be identified, and we directly set the predicted values of these nodes to be zero in the test phase.
\begin{algorithm}[!htb]
\footnotesize\caption{Training algorithm for \model}
\begin{algorithmic}[1]\label{train_algo}
\REQUIRE    %input
Encoder parameters $\Theta_{ENC}=(W_0,W_1,U_1,W_2,U_2,W_3,U_3)$, Decoder parameters $\Theta_{DEC}=(W_4, W_5)$
\ENSURE %output
Trained Model M
\FOR {each episode}
\STATE Draw network $G$ from distribution D (like the power-law model)
\STATE Calculate each node's exact BC value $b_v, \forall v \in V$
\STATE Get each node's embedding $z_v, \forall v \in V$ with Algorithm \ref{graphEmbed}
\STATE Compute BC ranking score $y_v$ for each node $v$ with Eq. (\ref{decoder})
\STATE Sample source nodes and target nodes, and form a batch of node pairs
\STATE Update $\Theta=(\Theta_{ENC}, \Theta_{DEC})$ with Adam by minimizing Eq. (\ref{loss})
\ENDFOR
\end{algorithmic}
\end{algorithm}

\subsection{Complexity Analysis}\label{sec:complexity}
\begin{figure}[t!]
\centering
\includegraphics[width=1.0\linewidth]{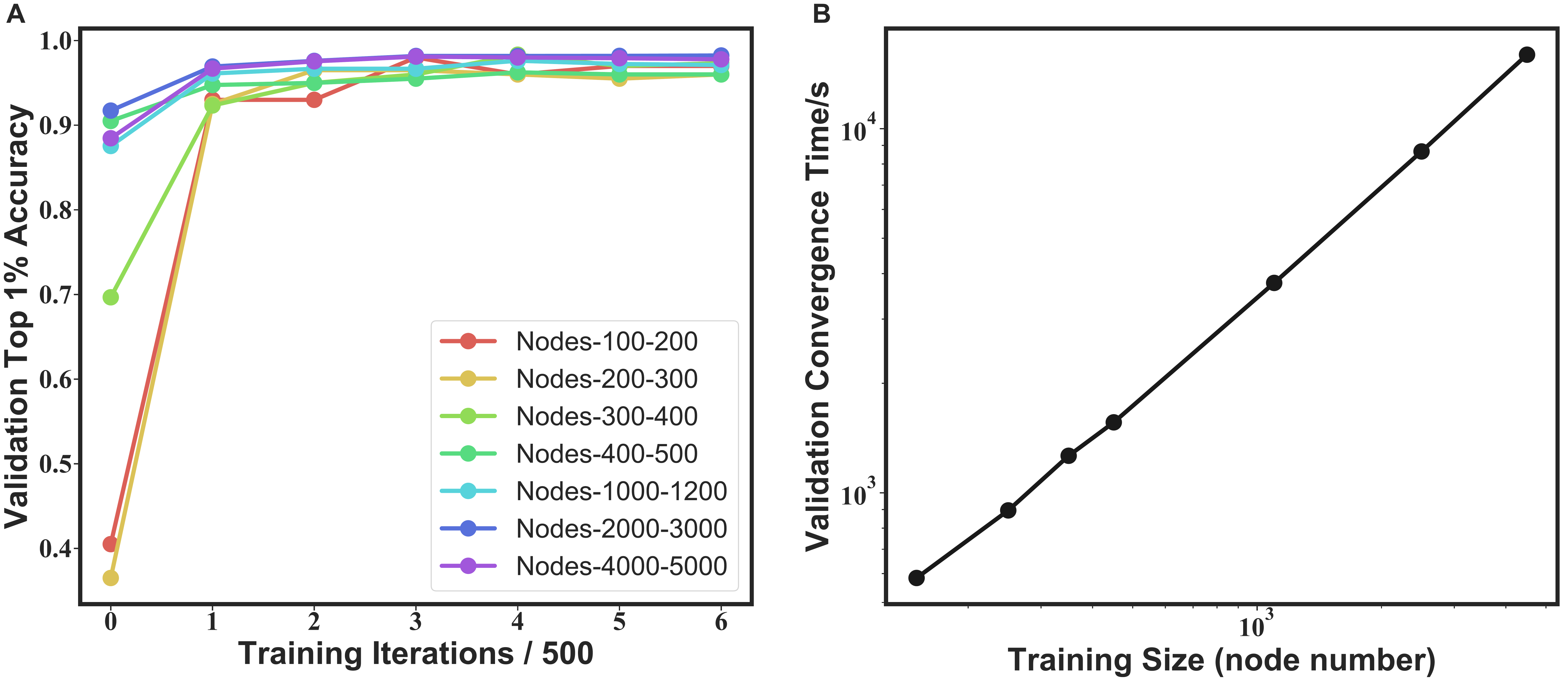}
\caption{Training analysis of \model. (A). \model~ convergence measured by validation top-1\% accuracy. Different lines denote different models trained with the corresponding scale. (B). \model~ convergence time for different training scales. We report all the time when training iterations reach 2500. The training size is the same as shown in (A).}\label{train_analysis}
\end{figure}

%In this section, we analyze the complexity of \model~for training and inference.
\noindent \textbf{Training complexity.}
During training, the time complexity is proportional to the training iterations, which are hard to be theoretically analyzed. Empirically, our model can converge (measured by validation performance) very quickly, as is shown in Figure \ref{train_analysis}(A), and the convergence time seems to increase linearly with the training scale (Figure \ref{train_analysis}(B)). It is generally affordable to train the \model. For example, to train a model with a training scale of 4000-5000, which we utilize in the following experiments, its convergence time is about 4.5 hours (Figure \ref{train_analysis}(B), including the time to compute ground truth BC values for training graphs already). Notably, the training phase is performed only once here, and we can utilize the trained model for any input network in the application phase.
%Note that the most time-consuming part in the training stage is to calculate exact BC values for training graphs, if we pre-calculate them before training starts, the illustrated time will be reduced greatly.

\noindent \textbf{Inference complexity.}
In the application phase, we apply the trained \model~ for a given network. The time complexity of the inference process is determined by two parts. The first is from the encoding phase. As shown in Algorithm \ref{graphEmbed}, the encoding complexity takes $O(L|V|N(\cdot))$, where $L$ is the number of propagation steps, and usually is a small constant (e.g., 5), $V$ is the number of nodes, $N(\cdot)$ is the average number of node neighbors. In practice, we use the adjacency matrix multiplication for Line 4-8 in Algorithm \ref{graphEmbed}, and due to the fact that most real-world networks are sparsely connected, the adjacency matrix in our setting is processed as a sparse matrix. As such, the encoding complexity in our implementations turns to be $O(|E|)$, where $E$ is the number of edges. The second is from the decoding phase. Once the nodes are encoded, we can compute their respective BC ranking score, and return the top-k highest BC nodes. the time complexity of this process mainly comes from the sorting operation, which takes $O(|V|)$. Therefore, the total time complexity for the inference phase should be $O(|E| + |V| + |V|log|V|)$.
% Due to that most real-world networks are sparsely connected, the adjacency matrix in our setting is processed as a sparse matrix. Hence, the total inference time complexity of \model~is $O(L|E|)$, where $L$ is the number of embedding propagations, and $|E|$ is the number of edges. 

%And the space complexity is $O(|V|p)$, where $|V|$ is the number of nodes, $p$ is the embedding dimension.
\section{Experiments}
In this section, we demonstrate the effectiveness and efficiency of \model~on both synthetic and real-world networks. We start by illustrating the discriminative embeddings generated by different models. %on a case of
%with a case study of synthetic network through 2D PCA projection. 
Then we explain experimental settings in detail, including baselines and datasets. After that, we discuss the results.

\subsection{Case Study: Visualization of Embeddings}
\begin{figure*}
\centering
\includegraphics[width=1.0\textwidth]{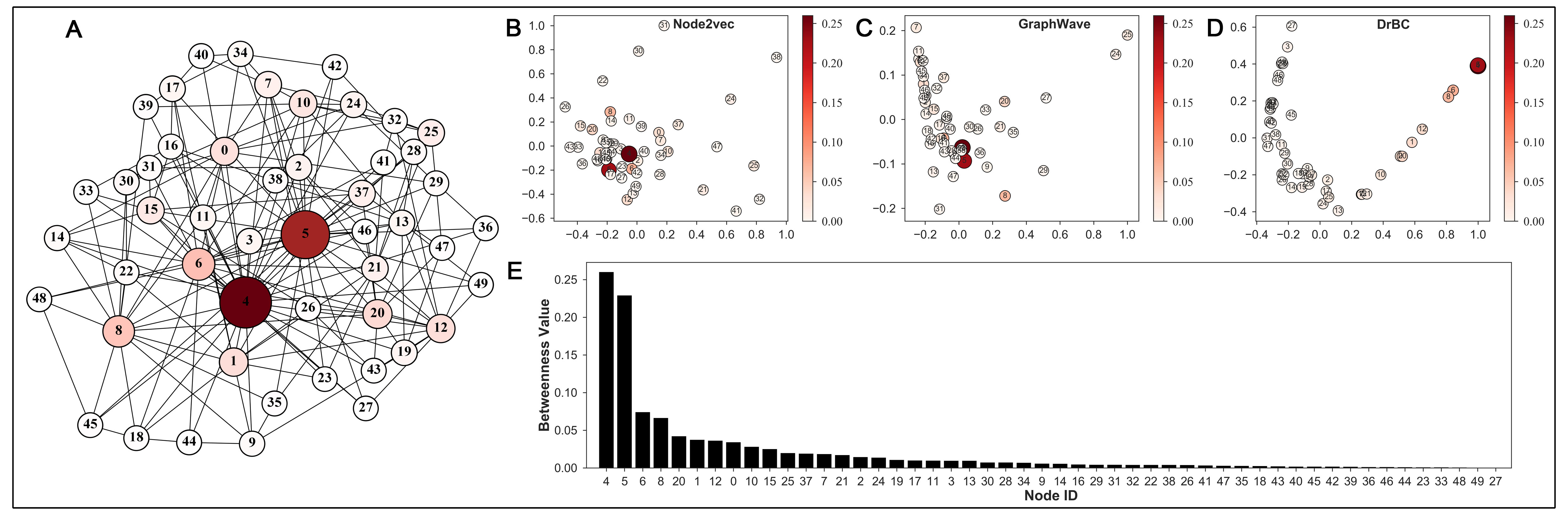}
\caption{The case network with 50 nodes and 200 edges, nodes with larger BC values having darker colors and larger shapes (A). 2D PCA projection of embeddings learned by Node2Vec (B), GraphWave (C) and \model~ (D). BC value distribution of the case network (descending order) (E).%\YS{When reproducing the figure, remember change the model name and enlarge the font of everything.}
}\label{demo}
\end{figure*}

We employ the 2D PCA projection to visualize the learned embeddings to intuitively show that our model preserves the relative BC order between nodes in the embedding space. For comparison, we also show the results of the other two traditional node embedding models, i.e., Node2Vec~\cite{grover2016node2vec} and GraphWave \cite{donnat2018learning}. Although these two models are not designed for BC approximation, they can be used to identify %nodes' equivalent structural roles
the equivalent structural roles in the network, and we believe nodes with similar BC values share similar ``bridge'' roles that control the information flow on networks. As such, we compare against these two models to see whether they can maintain BC similarity as well. The %case
example network is generated from the powerlaw-cluster model \cite{holme2002growing} (for which we use Networkx 1.11 and set the number of nodes $n=50$ and average degree $m=4$). For Node2Vec, let $p=1$, $q=2$ to enable it to capture the structural equivalence among nodes ~\cite{grover2016node2vec}. For GraphWave, we use the default settings in~\cite{donnat2018learning}.
As for \model, we use the same model tested in the following experiments. All embedding dimensions are set to be 128. As is shown in Figure~\ref{demo}, only in (D), the linear separable portions correspond to clusters of nodes with similar BC, while the other two both fail to make it. This indicates that \model~could generate more discriminative embeddings for BC prediction, which may provide some intuitions behind its prediction accuracy shown in the following experiments. 

\subsection{Experimental Setup} % and Evaluation Metric}
\subsubsection{\textbf{Baseline Methods.}}
We %compare
compare \model~with three approximation algorithms which focus on estimating exact BC values, %in this paper, 
i.e., ABRA, RK, k-BC, one approximation which seeks to identify top-$N\%$ highest BC nodes, i.e., KADABRA, and one traditional node embedding model, Node2Vec. Details of these baselines are described as follows:

\begin{itemize}[leftmargin=1em]
\item \textbf{ABRA~\cite{riondato2018abra}.} ABRA keeps sampling node pairs until the desired level of accuracy is reached. We use the parallel implementation by %Riondato et al.
\citeauthor{riondato2018abra}~\cite{riondato2018abra} and~\citeauthor{staudt2016networkit}~\cite{staudt2016networkit}. We set the error tolerance $\lambda$ to 0.01 and the probability $\delta$ to 0.1, following the setting in \cite{riondato2018abra}.

\item \textbf{RK~\cite{riondato2016fast}.} RK determines the required sample size based on the diameter of the network. We adopt the implementation in the NetworKit library \cite{staudt2016networkit}. We set the %allowed
error tolerance $\lambda$ to 0.01 and the probability $\delta$ to 0.1.

\item \textbf{k-BC~\cite{pfeffer2012k}.} k-BC bounds the traversals of Brandes algorithm \cite{brandes2001faster} by $k$ steps. For the value of $k$, we set it to be $20\%$ of the diameter of the network.

\item \textbf{KADABRA~\cite{borassi2016kadabra}.} KADABRA follows the idea of adaptive sampling and proposes balanced bidirectional BFS to sample the shortest paths. We use its variant well designed for computing top-$N\%$ highest BC nodes. We set the error tolerance and probability the same as ABRA and RK.

\item \textbf{Node2Vec~\cite{grover2016node2vec}.} Node2Vec designs a biased random walk procedure to explore a node's diverse neighbors, and learns each node's embedding vector that maximizes the likelihood of preserving its neighbors. Node2Vec can efficiently learn task-independent representations which are highly related to the network structure. In our experiments, we set $p$=1 and $q$=2 to enable Node2Vec to capture the structural equivalence of nodes, since BC can be viewed as a measure of ``bridge`` role on networks. %To apply Node2Vec to large unseen networks with no BC values provided, we train a mapping model (MLPs here) from synthetic graphs,  which takes input the embeddings from Node2Vec, and outputs the scalars as BC ranking scores. The training procedure is similar as \model (Algorithm \ref{train_algo}), we use the pairwise ranking loss to update the model parameters. In the test stage, we first run Node2Vec on the network, and obtain each node's embeddings, then the trained model is utilized to transform the embeddings to scalars which are used for BC ranking.
We train a MLP to map Node2Vec embeddings to BC ranking scores, following the same training procedure as Algorithm \ref{train_algo}. At the test stage, this baseline generates Node2Vec embeddings and then applies the trained MLP to obtain the ranking score for each node. 

\end{itemize}

\subsubsection{\textbf{Datasets.}} We evaluate the performance of \model~on both synthetic networks and large real-world ones. For synthetic networks, we generate them with the powerlaw-cluster model~\cite{holme2002growing}, which generates graphs that could capture both powerlaw degree distribution and small-world phenomenon, and most real-world networks conform to these two properties. % we believe it is a better model to simulate real networks.% Table \ref{graph types effects} shows the model trained on this type of graphs have better generalizability than those trained on other types of synthetic graphs. 
The basic parameters of this model are average degree $m=4$ and the probability of adding a triangle after adding a random edge $p=0.05$. We keep them the same when generating synthetic graphs with six different scales: 5000, 10000, 20000, 50000 and 100000. %For each size, we generate 30 random networks, and use the averaged performance over them to evaluate the model. 
For real-world test data, we use five large real-world networks %collected
provided by \citeauthor{alghamdi2017benchmark}~\cite{alghamdi2017benchmark}. Descriptions of these networks are as follows and Table \ref{real-world graphs} summarizes their statistics.%\FC{May add some new dataset}

% \noindent \textbf{as-skitter} is an undirected network of autonomous systems on the Internet connected to each other from the Skitter project. Nodes in the network represent IPv4 addresses and edges represent the topological connections between nodes.
\noindent \textbf{com-Youtube} is a video-sharing web site that includes a social network. Nodes are users and edges are friendships. 

\noindent \textbf{Amazon} is a product network created by crawling the Amazon online store. Nodes represent products and edges link commonly co-purchased products.

\noindent \textbf{Dblp} is an authorship network %of
extracted from the DBLP computer science bibliography. Nodes are authors and publications. Each edge connects an author to one of his publications.

% \noindent \textbf{Wiki-Talk} contains all the users and discussions from the inception of Wikipedia till January 2008. Nodes in the network represent Wikipedia users and edges represent the relationship between users.
\noindent \textbf{cit-Patents} is a citation network of U.S. patents. Nodes are patents and edges represent citations. In our experiments, we regard it as an undirected network.

\noindent \textbf{com-lj} is a social network where nodes are LiveJournal users and edges are their friendships.

% \noindent \textbf{livejournal-links} is the social network of LiveJournal users and their connections.

\begin{table}[t!]
\caption{\textbf{Summary of real-world datasets.}}\label{real-world graphs}
\resizebox{\linewidth}{!}{%
\begin{tabular}{@{}|c|c|c|c|c|@{}}
\toprule
Network         & |V|       & |E|        & Average Degree & Diameter \\ \midrule
com-Youtube     & 1,134,890 & 2,987,624 & 5.27          & 20       \\
\hline
Amazon        & 2,146,057 & 5,743,146  & 5.35           & 28       \\
\hline
Dblp     & 4,000,148 & 8,649,011  & 4.32           & 50        \\ 
\hline
cit-Patents   & 3,764,117 & 16,511,741  & 8.77           & 26       \\
\hline
com-lj & 3,997,962 & 34,681,189 & 17.35         & 17
\\ \bottomrule
% livejournal-links & 5,189,809 & 49,151,786 & 18.94  & 23  \\ \bottomrule
\end{tabular}}%
\end{table}

\subsubsection{\textbf{Ground Truth Computation.}}
We exploit the graph-tool library~\cite{peixoto2014graph} to \nop{efficiently} compute the exact BC values for synthetic networks. For real-world networks, we use the exact BC values reported by \citeauthor{alghamdi2017benchmark}~\cite{alghamdi2017benchmark}, which are computed via a parallel implementation of Brandes algorithm using a 96,000-core supercomputer. 

\subsubsection{\textbf{Evaluation Metrics.}}
For all baseline methods and \model, we report their effectiveness in terms of top-$N\%$ accuracy and kendall tau distance, and their efficiency in terms of wall-clock running time. 
%Note that \model\ is trained to rank all the nodes, while we only focus on identifying high BC nodes, we report the kendall tau distance, which is a global ranking quality metric, in Appendix \ref{appendix_kendall tau}.

\noindent \textbf{Top-$N\%$ accuracy} is defined as the percentage of overlap between the top-$N\%$ nodes as returned by an approximation method and the top-$N\%$ nodes as identified by Brandes algorithm (ground truth):
 \begin{equation*}
     \text{Top-}N\% = \frac{|\{\text{returned top-}N\% \text{nodes}\} \cap \{\text{true top-}N\% \text{nodes}\}|}{\lceil |V|\times N\% \rceil}
 \end{equation*}
where $|V|$ is the number of nodes, and $\lceil x \rceil$ is the ceiling function. In our paper, we mainly compare top-$1\%$, top-$5\%$ and top-$10\%$.

\noindent \textbf{Kendall tau distance} is a metric that calculates the number of disagreements between the rankings of the compared methods.
\begin{equation*}
    K(\tau_1, \tau_2) = \frac{2(\alpha - \beta)}{n*(n-1)}
\end{equation*}
where $\alpha$ is the number of concordant pairs, and $\beta$ is the number of discordant pairs. The value of kendall tall distance is in the range [-1, 1], where 1 means that two rankings are in total agreement, while -1 means that the two rankings are in complete disagreement.
%  \begin{equation*}
%      \begin{aligned}
%       K(\tau_1, \tau_2) = |\{(i,j):i<j, (\tau_1(i)<\tau_1(j) \wedge \tau_2(i)>\tau_2(j)) \vee \\ 
%       (\tau_1(i)>\tau_1(j) \wedge \tau_2(i)<\tau_2(j))\}|
%     \end{aligned}
%  \end{equation*}
% where $\tau_1(i)$ and $\tau_2(i)$ are the rankings of the element $i$ in two ranking list $\tau_1$ and $\tau_2$ respectively. $K(\tau_1,\tau_2)$ will be 0 if two lists are identical and $n(n-1)/2$ ($n$ is the list size) if one list is the reverse of the other. Often, $K(\tau_1,\tau_2)$ is normalized by $n(n-1)/2$ to make the values lie in [0,1]. In this paper, we adopt the normalized kendall tau distance.
\noindent \textbf{Wall-clock running time} is defined as the actual time taken from the start of a computer program to the end, usually in seconds.
%... \YS{please describe it.}
\nop{We also report the wall-clock time\nop{execution time} for each method to measure its efficiency. }

\subsubsection{\textbf{Other Settings.}}
All the experiments are conducted on a 80-core server with 512GB memory, and 8 16GB Tesla V100 GPUs. Notably, we train \model~with the GPUs while test it with only CPUs for a more fair time comparison with the baselines, since most baselines do not utilize the GPU environment. We generate 10,000 synthetic networks at random for training, and 100 for validation. We adopt early stopping to choose the best model based on validation performance. The model is implemented in Tensorflow with the Adam optimizer, and values of hyper-parameters (Table \ref{hyper-param}) are determined according to the performance on the validation set. %We will later release our code and trained models to support further research for this direction.\FC{Or should we release them now?}
The trained model and the implementation codes are released at \url{https://github.com/FFrankyy/DrBC}.

\begin{table}[t!]
\caption{Hyper-parameter configuration for \model.}\label{hyper-param}
\resizebox{\linewidth}{!}{%
\begin{tabular}{|c|c|l|}
\toprule
\multicolumn{1}{|c|}{\textbf{Hyper-parameter}} & \textbf{Value} & \multicolumn{1}{c|}{\textbf{Description}}                                                                                          \\ \midrule
learning rate                               & 0.0001         & the learning rate used by Adam optimizer                                                                                          \\
\hline
embedding dimension                               & 128            & dimension of node embedding vector                                                                                                \\
\hline
mini-batch size                                 & 16             & size of mini-batch training samples                                                                                                    \\
\hline
average node sampling times             & 5             & average sampling times per node for training                                                                                                    \\
\hline
maximum episodes                             & 10000          & maximum episodes for the training process \\
\hline
layer iterations                               & 5              & number of neighborhood-aggregation iterations                                                                                     \\ \bottomrule
\end{tabular}}%
\end{table}

\begin{table*}[t!]
\caption{Top-$N\%$ accuracy ($\times 0.01$) on synthetic graphs of different scales. The bold results indicate the best among all methods. For each scale, we report the mean and standard deviation over 30 tests.} \label{syn_test}
\resizebox{\linewidth}{!}{%
\begin{tabular}{@{}|cIc|c|c|c|c|cIc|c|c|c|c|cIc|c|c|c|c|c|@{}}
\toprule
\multirow{2}{*}{Scale}   & \multicolumn{6}{cI}{Top-$1\%$}     & \multicolumn{6}{cI}{Top-$5\%$}     & \multicolumn{6}{c|}{Top-$10\%$} \\ \cmidrule(l){2-19} 
                           & ABRA  & RK &k-BC & KADABRA & Node2Vec  & \model   & ABRA  & RK &k-BC  & KADABRA & Node2Vec & \model  & ABRA  & RK &k-BC & KADABRA & Node2Vec & \model \\ \midrule
5000        &\pmb{97.8$\pm$1.5}  &96.8$\pm$1.7  &94.1$\pm$0.8  &76.2$\pm$12.5  &19.1$\pm$4.8  &96.5$\pm$1.8
            &\pmb{96.9$\pm$0.7}  &95.6$\pm$0.9  &89.3$\pm$3.9  &68.7$\pm$13.4  &23.3$\pm$3.6  &95.9$\pm$0.9
            &\pmb{96.1$\pm$0.7}  &94.3$\pm$0.9  &86.7$\pm$4.5  &67.2$\pm$12.5  &25.4$\pm$3.4  &94.8$\pm$0.7 \\ \midrule
            
10000       &\pmb{97.2$\pm$1.2}  &96.4$\pm$1.3  &93.3$\pm$3.1  &74.6$\pm$16.5  &21.2$\pm$4.3  &96.7$\pm$1.2
            &\pmb{95.6$\pm$0.8}  &94.1$\pm$0.8  &88.4$\pm$5.1  &70.7$\pm$13.8  &20.5$\pm$2.7  &95.0$\pm$0.8
            &\pmb{94.1$\pm$0.6}  &92.2$\pm$0.9 &86.0$\pm$5.9  &67.8$\pm$13.0  &25.4$\pm$3.4  &94.0$\pm$0.9 \\ \midrule
            
20000       &\pmb{96.5$\pm$1.0}  &95.5$\pm$1.1  &91.6$\pm$4.0  &74.6$\pm$16.7  &16.1$\pm$3.9  &95.6$\pm$0.9
            &\pmb{93.9$\pm$0.8}  &92.2$\pm$0.9  &86.9$\pm$6.2  &69.1$\pm$13.5  &16.9$\pm$2.0  &93.0$\pm$1.1
            &\pmb{92.1$\pm$0.8}  &90.6$\pm$0.9  &84.5$\pm$6.8  &66.1$\pm$12.4  &19.9$\pm$1.9  &91.9$\pm$0.9 \\ \midrule
            
50000       &\pmb{94.6$\pm$0.7}  &93.3$\pm$0.9  &90.1$\pm$4.7  &73.8$\pm$14.9  &9.6$\pm$1.3  &92.5$\pm$1.2
            &\pmb{90.1$\pm$0.8}  &88.0$\pm$0.8  &84.4$\pm$7.2  &65.8$\pm$11.7  &13.8$\pm$1.0  &89.2$\pm$1.1
            &87.4$\pm$0.9  &\pmb{88.2$\pm$0.5}  &82.1$\pm$8.0  &61.3$\pm$10.4  &18.0$\pm$1.2  &87.9$\pm$1.0 \\ \midrule
            
100000      &\pmb{92.2$\pm$0.8}  &91.5$\pm$0.8  &88.6$\pm$4.7  &67.0$\pm$12.4  &9.6$\pm$1.3  &90.3$\pm$0.9
            &85.6$\pm$1.1  &\pmb{87.6$\pm$0.5}  &82.4$\pm$7.5  &57.0$\pm$9.4  &12.9$\pm$1.2  &86.2$\pm$0.9
            &81.8$\pm$1.5  &\pmb{87.4$\pm$0.4}  &80.1$\pm$8.2  &52.4$\pm$8.2  &17.3$\pm$1.3  &85.0$\pm$0.9 \\ \bottomrule
\end{tabular}}
\end{table*}

% \begin{table}[H]
% \caption{Kendall tau distance (\%) on synthetic graphs. (Since KADABRA only outputs the top-$N\%$ nodes, we cannot obtain the kendall tau distance for the overall ranking list.)}\label{syn_kendall tau}
% \resizebox{\linewidth}{!}{%
% \begin{tabular}{|c|c|c|c|c|c|}
% \toprule
% \diagbox{Method}{Accuracy}{Scale} & 5000 & 10000 & 20000 & 50000 & 100000 \\ \midrule
% ABRA  & 86.6$\pm$1.0 & 81.6$\pm$1.2 & 76.9$\pm$1.5 & 68.2$\pm$1.3 & 60.3$\pm$1.9 \\ \hline
% RK  & 78.6$\pm$0.6 & 72.3$\pm$0.6 & 65.5$\pm$1.2 & 53.3$\pm$1.4 & 44.2$\pm$0.2 \\ \hline
% k-BC  & 66.2$\pm$11.4 & 67.2$\pm$13.5 & 67.1$\pm$14.3 & 66.2$\pm$14.1 & 64.9$\pm$13.5 \\ \hline
% KADABRA  & NA & NA & NA & NA & NA \\ \hline 
% Node2Vec  &11.3$\pm$3.0  &8.5$\pm$2.3 &7.5$\pm$2.2  &7.1$\pm$1.8   &7.1$\pm$1.9   \\ \hline
% \model  & \pmb{88.4$\pm$0.3} & \pmb{86.8$\pm$0.4} & \pmb{84.0$\pm$0.5} & \pmb{80.1$\pm$0.5} & \pmb{77.8$\pm$0.4} \\ \bottomrule
% \end{tabular}}
% \end{table}

\begin{table}[t!]
\caption{Kendall tau distance ($\times 0.01$) on synthetic graphs. (Since KADABRA only outputs the top-$N\%$ nodes, we cannot obtain the kendall tau distance for the overall ranking list.)}\label{syn_kendall tau}
\resizebox{\linewidth}{!}{%
\begin{tabular}{|c|c|c|c|c|c|c|}
\toprule
\diagbox{Scale}{Kendal}{Method} & ABRA & RK & k-BC & KADABRA & Node2Vec & \model\\ \midrule
5000  & 86.6$\pm$1.0 & 78.6$\pm$0.6 & 66.2$\pm$11.4 & NA &11.3$\pm$3.0 & \pmb{88.4$\pm$0.3} \\ \hline
10000 & 81.6$\pm$1.2 & 72.3$\pm$0.6 & 67.2$\pm$13.5 & NA &8.5$\pm$2.3 & \pmb{86.8$\pm$0.4}\\ \hline
20000  & 76.9$\pm$1.5 & 65.5$\pm$1.2 & 67.1$\pm$14.3 & NA &7.5$\pm$2.2 & \pmb{84.0$\pm$0.5}\\ \hline
50000   & 68.2$\pm$1.3 & 53.3$\pm$1.4 & 66.2$\pm$14.1 & NA &7.1$\pm$1.8 & \pmb{80.1$\pm$0.5} \\ \hline 
100000  & 60.3$\pm$1.9 & 44.2$\pm$0.2 & 64.9$\pm$13.5 & NA &7.1$\pm$1.9 & \pmb{77.8$\pm$0.4}  \\ \bottomrule
\end{tabular}}
\end{table}

\subsection{Results on Synthetic Networks} \label{sec:result_syn}

We report the top-$N\%$ (1,5,10) accuracy, kendall tau distance and running time of baselines and \model~on synthetic networks with different scales, as shown in Table~\ref{syn_test},~\ref{syn_kendall tau} and~\ref{syn_time}. For each scale, we generate 30 networks at random for testing and report the mean and standard deviation. For ABRA, RK and KADABRA, we independantly run 5 times each on all the test graphs. The \model~ model is trained on powerlaw-cluster graphs with node sizes in 4000-5000.

% \begin{table}[H]
% \caption{Running time comparison on synthetic networks.} \label{syn_time}
% \resizebox{\linewidth}{!}{%
% \begin{tabular}{|c|c|c|c|c|c|}
% \toprule
% \diagbox{Method}{Time/s}{Scale} & 5000 & 10000 & 20000 & 50000 & 100000 \\ \midrule
% ABRA & 18.5$\pm$3.6 & 29.2$\pm$4.8  & 52.7$\pm$8.1    & 168.3$\pm$23.8  & 380.3$\pm$63.7    \\ \hline
% RK  & 17.1$\pm$3.0 & 21.0$\pm$3.6   & 43.0$\pm$3.2    & 131.4$\pm$2.0   & 363.4$\pm$36.3   \\ \hline
% k-BC & 12.2$\pm$6.3 & 47.2$\pm$27.3 & 176.4$\pm$105.1 & 935.1$\pm$505.9 & 3069.2$\pm$1378.5 \\ \hline
% KADABRA   &0.6$\pm$0.1 	&1.0$\pm$0.2  	&1.6$\pm$0.3  	&3.9$\pm$1.0 	&\pmb{7.2$\pm$1.8}   \\ \hline
% Node2Vec &32.4$\pm$3.8   &73.1$\pm$7.0   &129.3$\pm$17.6  &263.2$\pm$46.6 &416.2$\pm$37.0    \\ \hline
% \model   & \pmb{0.3$\pm$0.0} & \pmb{0.6$\pm$0.0} & \pmb{1.4$\pm$0.0} & \pmb{3.9$\pm$0.2} & 8.2$\pm$0.3 \\ \bottomrule
% \end{tabular}}
% \end{table}

\begin{table}[t!]
\caption{Running time comparison on synthetic networks.} \label{syn_time}
\resizebox{\linewidth}{!}{%
\begin{tabular}{|c|c|c|c|c|c|c|}
\toprule
\diagbox{Scale}{Time/s}{Method} & ABRA & RK & k-BC & KADABRA & Node2Vec & \model\\ \midrule
5000    & 18.5$\pm$3.6 & 17.1$\pm$3.0 & 12.2$\pm$6.3  &0.6$\pm$0.1 	&32.4$\pm$3.8  & \pmb{0.3$\pm$0.0}     \\ \hline
10000  & 29.2$\pm$4.8 & 21.0$\pm$3.6 & 47.2$\pm$27.3 &1.0$\pm$0.2 &73.1$\pm$7.0 & \pmb{0.6$\pm$0.0}\\ \hline
20000   & 52.7$\pm$8.1 & 43.0$\pm$3.2 & 176.4$\pm$105.1 &1.6$\pm$0.3 &129.3$\pm$17.6 & \pmb{1.4$\pm$0.0}\\ \hline
50000   & 168.3$\pm$23.8  & 131.4$\pm$2.0 & 935.1$\pm$505.9 &3.9$\pm$1.0  &263.2$\pm$46.6 & \pmb{3.9$\pm$0.2}\\ \hline 
100000  & 380.3$\pm$63.7 & 363.4$\pm$36.3 & 3069.2$\pm$1378.5 &\pmb{7.2$\pm$1.8}  &416.2$\pm$37.0 & 8.2$\pm$0.3\\ \bottomrule
\end{tabular}}
\end{table}

\begin{table}[t!]
\caption{\model's generalization results on different scales (Top-$1\%$ accuracy, $\times 0.01$).
} \label{syn_generalization_top1}
\resizebox{\linewidth}{!}{%
\begin{tabular}{|c|c|c|c|c|c|}
\toprule
\diagbox{Train}{Accuracy}{Test}  & 5000 & 10000 & 20000 & 50000 & 100000 \\ \midrule
100\_200  & 90.5$\pm$2.9 & 88.3$\pm$2.1 & 85.5$\pm$1.9 & 83.9$\pm$1.1 & 82.2$\pm$0.9 \\ \hline
200\_300  & 92.5$\pm$2.7 & 90.0$\pm$2.2 & 87.0$\pm$2.1 & 84.8$\pm$1.1 & 82.9$\pm$0.9 \\ \hline
1000\_1200  & 94.3$\pm$2.2 & 90.6$\pm$1.7 & 87.8$\pm$1.9 & 85.1$\pm$1.1 & 83.1$\pm$0.9 \\ \hline
2000\_3000  & 95.7$\pm$1.8 & 93.5$\pm$1.7 & 90.7$\pm$1.6 & 87.8$\pm$1.1 & 85.9$\pm$0.7 \\ \hline
4000\_5000  & \pmb{96.5$\pm$1.8} & \pmb{96.7$\pm$1.2} & \pmb{95.6$\pm$0.9} & \pmb{92.5$\pm$1.2} & \pmb{90.3$\pm$0.9} \\ \bottomrule
\end{tabular}}
\end{table}

\begin{table}[t!]
\caption{\model's generalization results on different scales (kendall tau distance, $\times 0.01$).
} \label{syn_generalization_kendall tau}
\resizebox{\linewidth}{!}{%
\begin{tabular}{|c|c|c|c|c|c|}
\toprule
\diagbox{Train}{Kendal}{Test}  & 5000 & 10000 & 20000 & 50000 & 100000 \\ \midrule
100\_200  & 43.6$\pm$1.0 & 40.1$\pm$0.6 & 37.5$\pm$0.5 & 35.2$\pm$0.3 & 33.9$\pm$0.2 \\ \hline
200\_300  & 42.7$\pm$0.8 & 39.5$\pm$0.5 & 37.1$\pm$0.5 & 35.0$\pm$0.3 & 33.9$\pm$0.2 \\ \hline
1000\_1200  & 56.4$\pm$0.6 & 42.7$\pm$0.4 & 36.0$\pm$0.4 & 31.7$\pm$0.3 & 29.8$\pm$0.2 \\ \hline
2000\_3000  & 86.1$\pm$0.3 & 78.1$\pm$0.5 & 69.9$\pm$0.6 & 62.3$\pm$0.5 & 59.1$\pm$0.3 \\ \hline
4000\_5000  & \pmb{88.4$\pm$0.3} & \pmb{86.8$\pm$0.4} & \pmb{84.0$\pm$0.5} & \pmb{80.1$\pm$0.5} & \pmb{77.8$\pm$0.4} \\ \bottomrule
\end{tabular}}
\end{table}

\begin{table*}[t!]
\caption{Top-$N\%$ accuracy ($\times 0.01$) and running time on large real-world networks. *result is adopted from \cite{alghamdi2017benchmark}, since RK can not finish within the acceptable time. The bold results indicate the best performance of the network under the current metric.} \label{real_test}
\resizebox{\linewidth}{!}{%
\begin{tabular}{@{}|cIc|c|c|c|cIc|c|c|c|cIc|c|c|c|cIc|c|c|c|c|@{}}
\toprule
\multirow{2}{*}{Network}   & \multicolumn{5}{cI}{Top-$1\%$}     & \multicolumn{5}{cI}{Top-$5\%$}     & \multicolumn{5}{cI}{Top-$10\%$}      & \multicolumn{5}{c|}{Time/s}          \\ \cmidrule(l){2-21} 
                           & ABRA  & RK  & KADABRA & Node2Vec  & \model   & ABRA  & RK  & KADABRA & Node2Vec & \model  & ABRA  & RK  & KADABRA & Node2Vec & \model  & ABRA    & RK     & KADABRA & Node2Vec & \model \\ \midrule
com-youtube                &\pmb{95.7}       & 76.0 & 57.5 &12.3      & 73.6    &\pmb{91.2}       & 75.8 & 47.3 &18.9      & 66.7     &89.5       & \pmb{100.0} & 44.6    &23.6   & 69.5     &72898.7       & 125651.2 & \pmb{116.1}  &4729.8  & 402.9   \\ \midrule
amazon                     & 69.2 & 86.0 & 47.6 &16.7      & \pmb{86.2}     & 58.0 & 59.4 & 56.0 &23.2      & \pmb{79.7}     & 60.3 & \pmb{100.0} & 56.7 &26.6      & 76.9     & 5402.3  & 149680.6 & \pmb{244.7} &10679.0     & 449.8   \\ \midrule
Dblp                       &49.7       & NA      & 35.2 &11.5      & \pmb{78.9}     &45.5       & NA      & 42.6 &20.2      & \pmb{72.0}    &\pmb{100.0}       & NA      & 50.4 &27.7      & 72.5     &11591.5         & NA         & \pmb{398.1} &17446.9    & 566.7   \\ \midrule
cit-Patents               & 37.0 & \pmb{74.4} & 23.4 &0.04      & 48.3    & 42.4 & \pmb{68.2} & 25.1 &0.29      & 57.5    & 50.9 & 53.5 & 21.6 &0.99      & \pmb{64.1}     & 10704.6 & 252028.5 & \pmb{568.0} &11729.1    & 744.1   \\ \midrule
com-lj            & 60.0 & 54.2* & 31.9 &3.9      & \pmb{67.2}    & 56.9 &NA       & 39.5 &10.35      & \pmb{72.6}    & 63.6 & NA      & 47.6 &15.4      & \pmb{74.8}     & 34309.6 & NA & \pmb{612.9} &18253.6  & 2274.2  \\ \bottomrule
\end{tabular}}
\end{table*}

We can see from Table~\ref{syn_test}, \ref{syn_kendall tau} and \ref{syn_time} that, \model~ achieves competitive top-$N\%$ accuracy compared with the best approximation results, and outperforms all baselines in terms of the kendall tau distance. Running time comparison shows the obvious efficiency advantage. Take graphs with size 100000 as an example, although \model~sacrifices about 2\% loss in top-$1\%$ accuracy compared with the best result (ABRA), it is over 46 times faster. It is noteworthy that we do not consider the training time here. Since none of the approximation algorithms requires training, it may cause some unfair comparison. However, considering that \model~only needs to be trained once offline and can then generalize to any unseen network, it is reasonable not to consider the training time in comparison. As analyzed in section \ref{sec:complexity}, \model~can converge rapidly, resulting in acceptable training time, e.g., it takes about 4.5 hours to train and select the model used in this paper. 

Overall, ABRA achieves the highest top-$N\%$ accuracy, and RK is very close to ABRA in accuracy and time, while k-BC is far worse in terms of both effectiveness and efficiency. KADABRA is designed to identify the top-$N\%$ highest BC nodes, so it only outputs the top-$N\%$ nodes, making it impossible to calculate the kendall tau distance for the whole ranking list. We can see from Table \ref{syn_time} that KADABRA is very efficient, which is close to \model, however, its high efficiency sacrifices too much accuracy, over 20\% lower than \model~on average. We also observe that Node2Vec performs very poor in this task, which may due to that Node2Vec learns task-independent representations which only capture nodes' local structural information, while BC is a measure highly related to the global. In our model, we train it in an end-to-end manner, enabling the ground truth BC relative order to shape the learned representations. Experiments demonstrate that learning in this way can capture better informative features related to BC. For the kendall tau distance metric which measures the quality of the whole ranking list, \model~performs significantly better than the other baselines. This is because \model~learns to maintain the relative order between nodes specified by true BC values, while other baselines either focus on approximating exact BC values or seek to identify the top-$N\%$ highest BC nodes.

The inductive setting of \model~enables us to train and test our model on networks of different scales, since the model's parameters are independent of the network scale. Table \ref{syn_test} and \ref{syn_kendall tau} have already verified that our model can generalize to larger graphs than what they are trained on. Here we show the model's full generalizability by training on different scales and compare the generalizability for each scale. As is shown in Table \ref{syn_generalization_top1} and \ref{syn_generalization_kendall tau} which illustrate results of top-$1\%$ accuracy and kendall tau distance, %(see Appendix \ref{appendix_full_generality_synthetic} for other metrics results)
the model can generalize well on larger graphs for each scale, and it seems that model trained on larger scales can achieve better generalization results. The intuition behind is that larger scale graphs represent more difficult samples, making the learned model more generalizable. %Note that it may be the reason of over-fitting that Model 4000-5000 performs worse than model 2000-3000 on graphs 3000. 
This observation may inspire us to train on larger scales to improve the performance on very large real-world networks. In this paper, we just use the model trained with node sizes in 4000-5000 and test on the following five large real-world networks.

\subsection{Results on Real-world Networks}\label{sec:real-world}
In section~\ref{sec:result_syn}, we test \model~on synthetic graphs generated from the same model as it is trained on, and we have observed it can generalize well on those larger than the training scale. In this section, we test \model~on different large-scale real-world networks to see whether it is still generalizable in practice. 

We compare top-$N\%$ accuracy, kendall tau distance and running time in Table \ref{real_test} and \ref{real_kendall tau}. Since k-BC cannot obtain results within the acceptable time on these networks, we do not compare against it here. RK cannot finish within 3 days for Dblp and com-lj, so we just use its top-$1\%$ accuracy on com-lj reported in \cite{alghamdi2017benchmark}. Due to the time limits, for ABRA and RK, we only run once on each network. For KADABRA and \model, we independently run five times for each network and report the averaged accuracy and time.

% \begin{table}[H]
% \caption{Kendall tau distance (\%) on real-world networks. (Since KADABRA only outputs top-$N\%$ nodes, we cannot obtain the kendall tau distance for the overall ranking list. RK did not finish on Dblp and com-lj within acceptable time.)}\label{real_kendall tau}
% \resizebox{\linewidth}{!}{%
% \begin{tabular}{|c|c|c|c|c|c|}
% \toprule
% \diagbox{Method}{kendall tau}{Network}& com-youtube & amazon & Dblp & cit-Patents & com-lj \\ \midrule
% ABRA    &56.2             & 16.3 & 14.3      & 17.3     & 22.8 \\ \hline
% RK    & 13.9      &9.7 &NA        & 15.3      &   NA \\ \hline
% KADABRA & NA          & NA     & NA     & NA          & NA     \\ \hline
% Node2Vec &46.2           &44.7       &49.5       &4.0            &35.1       \\ \hline
% \model  &\pmb{57.3}      & \pmb{69.3} & \pmb{71.9} & \pmb{72.6}      & \pmb{71.3}  \\ \bottomrule
% \end{tabular}}
% \end{table}

\begin{table}[t!]
\caption{Kendall tau distance ($\times 0.01$) on real-world networks. (Since KADABRA only outputs top-$N\%$ nodes, we cannot obtain the kendall tau distance for the overall ranking list. RK did not finish on Dblp and com-lj within the acceptable time.)}\label{real_kendall tau}
\resizebox{\linewidth}{!}{%
\begin{tabular}{|c|c|c|c|c|c|}
\toprule
\diagbox{Network}{kendall tau}{Method} &ABRA &RK &KADABRA &Node2Vec &\model \\ \midrule
com-youtube   &56.2 & 13.9 & NA &46.2 &\pmb{57.3}            \\ \hline
amazon     & 16.3 &9.7  & NA &44.7 & \pmb{69.3} \\ \hline
Dblp  & 14.3 &NA & NA &49.5  & \pmb{71.9}\\ \hline
cit-Patents  & 17.3 & 15.3  & NA &4.0   & \pmb{72.6}\\ \hline
com-lj   & 22.8 &NA & NA &35.1 & \pmb{71.3}\\ \bottomrule
\end{tabular}}
\end{table}

We can see in Table \ref{real_test} that different methods perform with a large variance on different networks. Take ABRA as an example, it can achieve 95.7\% top 1\% accuracy on com-youtube, while only 37.0\% on cit-Patents. It seems that no method can achieve consistent overwhelming accuracy than others. However, if we consider the trade-off between accuracy and efficiency, our model performs the best, especially for the latter, which is several orders of magnitude faster than ABRA and RK. Besides efficiency, \model~actually performs on par with or better than these approximation algorithms in terms of accuracy. Specifically, \model~ achieves the best top-$1\%$ and top-$5\%$ accuracy in three out of the five networks. KADABRA is the most efficient one, while accompanied by too much accuracy loss. Node2Vec performs the worst, with the worst accuracy and relatively longer running time. In terms of the kendall tau distance, \model~consistently performs the best among all methods (Table \ref{real_kendall tau}).

\begin{table}[t!]
    \caption{Effect of different training graph types on synthetic graphs (\%). For each type, node sizes of the training graphs and test graphs both lie between 100-200. We report the mean and standard deviations over 100 tests.}\label{graph types synth}
    {\small
    \begin{tabular}{@{}|c|c|c|c|c|@{}}
    \toprule
    \diagbox{Train}{Top-$1\%$}{Test}     & ER     & BA     & PL-cluster \\ \hline
    ER          & \pmb{87.0$\pm$33.6} & 82.0$\pm$38.4 & 84.0$\pm$36.7     \\ \hline
    BA          & 62.5$\pm$48.2 & 93.0$\pm$25.5 & 93.0$\pm$25.5      \\ \hline
    PL-cluster &73.5$\pm$43.9  &\pmb{96.0$\pm$19.6}  &\pmb{96.0$\pm$19.6}       \\ \bottomrule
    \end{tabular}}%
    \vspace{-1em}
\end{table}

\begin{table}[t!]
\caption{Effect of different training graph types on real networks (\%).}\label{graph types real}
\resizebox{\linewidth}{!}{%
\begin{tabular}{|c|c|c|c|c|c|}
\toprule
\diagbox{Train}{Top-$1\%$}{Test}& com-youtube & amazon & Dblp & cit-Patents & com-lj \\ \midrule
ER    &66.58             & 73.88 & 64.82      & 38.51     & 54.26 \\ \hline
BA    & 72.92      &70.89 &73.87        & 35.56      &  55.08 \\ \hline
PL-cluster & \pmb{73.60}          &\pmb{86.15}     & \pmb{78.92}     & \pmb{48.31}         & \pmb{67.17}    \\ \hline
\end{tabular}}
\end{table}

\section{Discussion}
In this section, we discuss the potential reasons behind \model's success. Since it's hard to prove GNN's theoretical approximation bound, we explain by giving the following observations:
%why our model can achieve consistently promising performance on both synthetic graphs with varying sizes and real-world networks across different types. Although we do not provide theoretical approximation bound, we believe the performance is reasonable based on the following observations:

(1) The exact BC computation method, i.e. Brandes algorithm, inherently follows a similar neighbor aggregation schema (Eq. (\ref{update})) as our encoder, despite that their neighbors are on the shortest paths. We believe this common trait enables our model to capture characteristics that are essential to BC computation;
    %We believe our neighbor-aggregation encoder framework could kind of capture this 

(2) Our model limits graph exploration to a neighborhood of $L$-hops around each node, where $L$ denotes iterations of neighbor aggregation. The idea of using the $L$-hop sub-structure to approximate exact BC values is the basis of many existing BC approximation algorithms, such as EGO~\cite{everett2005ego} and $\kappa$-path~\cite{kourtellis2013identifying}. %$\kappa$-path has a lower bound guarantee, despite with some constraints. As for our model, there may also exist some theoretical guarantees, which we leave as a very important future direction;
    
(3) We train the model in an end-to-end manner with exact BC values as the ground truth. Like other successful deep learning applications on images or texts, if given enough training samples with ground truth, the model is expected to be able to learn well.
    %learn {\color{red}{some essences}}\muhao{what do some essences mean?} for the particular task, 
    %although nobody can explicitly tell the reason behind this kind of black models.
    % This is the similar case for our model, %we may not explain the exact reasons why it works, the model indeed automatically learns to capture the node similarities specified by their true BC values, 
    % as demonstrated in Figure \ref{demo}.
    %relative ranking orders between node pairs specified by their true BC values in a data-driven way. Visualizations of node embeddings learned by our model (Figure \ref{demo}) also indicate that the model captures the 

(4) The model is trained on synthetic graphs from the powerlaw-cluster (PL-cluster) model, which possesses the characteristics of power-law degree distribution and small-world phenomenon (large clustering coefficient), both of which appear in most real-world networks. In Table \ref{graph types synth}, we train \model~on Erd\H{o}s-R\'{e}nyi (ER)~\cite{erdds1959random}, Barab\'{a}si-Albert (BA)~\cite{barabasi1999emergence} and PL-cluster~\cite{holme2002growing} graphs, and test each on different types. The results show that \model~always performs the best when the training type is the same as the testing type, and the PL-cluster training type enables better generalizations for \model~over the other two types. Table \ref{graph types real} further confirms this conclusion on real-world networks.

% we show that models trained on this type of graphs have more powerful generalizability than those trained on other types of graphs.   %This may be reason behind that our model could generalize well across different types of large real-world networks.  

\section{Conclusion and Future Work}
In this paper, we for the first time investigate the power of graph neural networks in identifying high betweenness centrality nodes, which is a traditional but significant problem essential to many applications but often computationally prohibitive on large networks with millions of nodes.
%approximating a global network structural measure, the betweenness centrality (BC), which is essential to many applications but computationally prohibitive for large networks, . We design a simple but effective framework for BC approximation, %with
Our model is constituted with a neighborhood-aggregation encoder and a multi-layer perceptron decoder, and naturally applies to inductive settings where training and testing are independent. Extensive experiments on both synthetic graphs and large real-world networks show that our model can achieve very competitive accuracy while possessing a huge advantage in terms of running time. Notably, the presented results also highlight the importance of classical network models, such as the powerlaw-cluster model. Though extremely simple, it captures the key features, i.e., degree heterogeneity and small-world phenomenon, of many real-world networks. This tends out to be extremely important to train a deep learning model to solve very challenging problems on complex real networks.
%In the end, we
%We also apply \model\space to the network dismantling problem, and demonstrate its superior performance over other competing baselines on real-world datasets.

There are several future directions for this work, such as exploring theorectical guarantees, applying it to some BC-related downstream applications (e.g. community detection and network attack) and extending the framework to approximate other network structural measures.
%, such as the shortest distance between any two nodes. We believe the model's performance can be further improved by incorporating more techniques such as the utilization of sampling strategies from existing BC approximation algorithms, as well as attention-based neighborhood aggregation functions. Besides, we seek to apply our model to more downstream applications of BC, such as community detection which relies on the edge betweenness centrality. Moreover, we are also interested in extending our framework to approximate other network structural measures, such as the shortest distance between any two nodes, the closeness centrality~\cite{freeman1977set}, and the motif counts~\cite{benson2016higher}.

\begin{acks}
This work is partially supported by the China Scholarship Council (CSC),  NSF III-1705169, NSF CAREER Award 1741634, Amazon Research Award, and NSFC-71701205.
\end{acks}

\bibliographystyle{ACM-Reference-Format}
\bibliography{ref}


\begin{thebibliography}{37}

%%% the Web address.

\ifx \showCODEN    \undefined \def \showCODEN     #1{\unskip}     \fi
\ifx \showDOI      \undefined \def \showDOI       #1{#1}\fi
\ifx \showISBNx    \undefined \def \showISBNx     #1{\unskip}     \fi
\ifx \showISBNxiii \undefined \def \showISBNxiii  #1{\unskip}     \fi
\ifx \showISSN     \undefined \def \showISSN      #1{\unskip}     \fi
\ifx \showLCCN     \undefined \def \showLCCN      #1{\unskip}     \fi
\ifx \shownote     \undefined \def \shownote      #1{#1}          \fi
\ifx \showarticletitle \undefined \def \showarticletitle #1{#1}   \fi
\ifx \showURL      \undefined \def \showURL       {\relax}        \fi
% invisible to TeX
\providecommand\bibfield[2]{#2}
\providecommand\bibinfo[2]{#2}
\providecommand\natexlab[1]{#1}
\providecommand\showeprint[2][]{arXiv:#2}

\bibitem[\protect\citeauthoryear{AlGhamdi, Jamour, Skiadopoulos, and
  Kalnis}{AlGhamdi et~al\mbox{.}}{2017}]%
        {alghamdi2017benchmark}
\bibfield{author}{\bibinfo{person}{Ziyad AlGhamdi}, \bibinfo{person}{Fuad
  Jamour}, \bibinfo{person}{Spiros Skiadopoulos}, {and} \bibinfo{person}{Panos
  Kalnis}.} \bibinfo{year}{2017}\natexlab{}.
\newblock \showarticletitle{A benchmark for betweenness centrality
  approximation algorithms on large graphs}. In
  \bibinfo{booktitle}{\emph{SSDBM}}. \bibinfo{pages}{6}.
\newblock


\bibitem[\protect\citeauthoryear{Bai, Ding, Bian, Sun, and Wang}{Bai
  et~al\mbox{.}}{2018}]%
        {bai2018graph}
\bibfield{author}{\bibinfo{person}{Yunsheng Bai}, \bibinfo{person}{Hao Ding},
  \bibinfo{person}{Song Bian}, \bibinfo{person}{Yizhou Sun}, {and}
  \bibinfo{person}{Wei Wang}.} \bibinfo{year}{2018}\natexlab{}.
\newblock \showarticletitle{Graph Edit Distance Computation via Graph Neural
  Networks}.
\newblock \bibinfo{journal}{\emph{arXiv preprint arXiv:1808.05689}}
  (\bibinfo{year}{2018}).
\newblock


\bibitem[\protect\citeauthoryear{Barab{\'a}si and Albert}{Barab{\'a}si and
  Albert}{1999}]%
        {barabasi1999emergence}
\bibfield{author}{\bibinfo{person}{Albert-L{\'a}szl{\'o} Barab{\'a}si} {and}
  \bibinfo{person}{R{\'e}ka Albert}.} \bibinfo{year}{1999}\natexlab{}.
\newblock \showarticletitle{Emergence of scaling in random networks}.
\newblock \bibinfo{journal}{\emph{science}} \bibinfo{volume}{286},
  \bibinfo{number}{5439} (\bibinfo{year}{1999}), \bibinfo{pages}{509--512}.
\newblock


\bibitem[\protect\citeauthoryear{Borassi and Natale}{Borassi and
  Natale}{2016}]%
        {borassi2016kadabra}
\bibfield{author}{\bibinfo{person}{Michele Borassi} {and}
  \bibinfo{person}{Emanuele Natale}.} \bibinfo{year}{2016}\natexlab{}.
\newblock \showarticletitle{KADABRA is an ADaptive Algorithm for Betweenness
  via Random Approximation}. In \bibinfo{booktitle}{\emph{ESA}}.
\newblock


\bibitem[\protect\citeauthoryear{Brandes}{Brandes}{2001}]%
        {brandes2001faster}
\bibfield{author}{\bibinfo{person}{Ulrik Brandes}.}
  \bibinfo{year}{2001}\natexlab{}.
\newblock \showarticletitle{A faster algorithm for betweenness centrality}.
\newblock \bibinfo{journal}{\emph{Journal of mathematical sociology}}
  \bibinfo{volume}{25}, \bibinfo{number}{2} (\bibinfo{year}{2001}),
  \bibinfo{pages}{163--177}.
\newblock


\bibitem[\protect\citeauthoryear{Braunstein, Dall’Asta, Semerjian, and
  Zdeborov{\'a}}{Braunstein et~al\mbox{.}}{2016}]%
        {braunstein2016network}
\bibfield{author}{\bibinfo{person}{Alfredo Braunstein}, \bibinfo{person}{Luca
  Dall’Asta}, \bibinfo{person}{Guilhem Semerjian}, {and}
  \bibinfo{person}{Lenka Zdeborov{\'a}}.} \bibinfo{year}{2016}\natexlab{}.
\newblock \showarticletitle{Network dismantling}.
\newblock \bibinfo{journal}{\emph{PNAS}} \bibinfo{volume}{113},
  \bibinfo{number}{44} (\bibinfo{year}{2016}).
\newblock


\bibitem[\protect\citeauthoryear{Chen, Sun, Tian, et~al\mbox{.}}{Chen
  et~al\mbox{.}}{2018}]%
        {chen2018enhanced}
\bibfield{author}{\bibinfo{person}{Haochen Chen}, \bibinfo{person}{Xiaofei
  Sun}, \bibinfo{person}{Yingtao Tian}, {et~al\mbox{.}}}
  \bibinfo{year}{2018}\natexlab{}.
\newblock \showarticletitle{Enhanced Network Embeddings via Exploiting Edge
  Labels}. In \bibinfo{booktitle}{\emph{CIKM}}.
\newblock


\bibitem[\protect\citeauthoryear{Chen and Sun}{Chen and Sun}{2017}]%
        {chen2017task}
\bibfield{author}{\bibinfo{person}{Ting Chen} {and} \bibinfo{person}{Yizhou
  Sun}.} \bibinfo{year}{2017}\natexlab{}.
\newblock \showarticletitle{Task-guided and path-augmented heterogeneous
  network embedding for author identification}. In
  \bibinfo{booktitle}{\emph{WSDM}}.
\newblock


\bibitem[\protect\citeauthoryear{Chung and Lee}{Chung and Lee}{2014}]%
        {chung2014finding}
\bibfield{author}{\bibinfo{person}{Chin-Wan Chung} {and}
  \bibinfo{person}{Min-joong Lee}.} \bibinfo{year}{2014}\natexlab{}.
\newblock \showarticletitle{Finding k-highest betweenness centrality vertices
  in graphs}. In \bibinfo{booktitle}{\emph{23rd International World Wide Web
  Conference}}. International World Wide Web Conference Committee,
  \bibinfo{pages}{339--340}.
\newblock


\bibitem[\protect\citeauthoryear{Donnat, Zitnik, Hallac, and Leskovec}{Donnat
  et~al\mbox{.}}{2018}]%
        {donnat2018learning}
\bibfield{author}{\bibinfo{person}{Claire Donnat}, \bibinfo{person}{Marinka
  Zitnik}, \bibinfo{person}{David Hallac}, {and} \bibinfo{person}{Jure
  Leskovec}.} \bibinfo{year}{2018}\natexlab{}.
\newblock \showarticletitle{Learning Structural Node Embeddings via Diffusion
  Wavelets}.
\newblock  (\bibinfo{year}{2018}).
\newblock


\bibitem[\protect\citeauthoryear{Erd\H{o}s and R\'{e}nyi}{Erd\H{o}s and
  R\'{e}nyi}{1959}]%
        {erdds1959random}
\bibfield{author}{\bibinfo{person}{P Erd\H{o}s} {and} \bibinfo{person}{A
  R\'{e}nyi}.} \bibinfo{year}{1959}\natexlab{}.
\newblock \showarticletitle{On random graphs I}.
\newblock \bibinfo{journal}{\emph{Publ. Math. Debrecen}}  \bibinfo{volume}{6}
  (\bibinfo{year}{1959}).
\newblock


\bibitem[\protect\citeauthoryear{Everett and Borgatti}{Everett and
  Borgatti}{2005}]%
        {everett2005ego}
\bibfield{author}{\bibinfo{person}{Martin Everett} {and}
  \bibinfo{person}{Stephen~P Borgatti}.} \bibinfo{year}{2005}\natexlab{}.
\newblock \showarticletitle{Ego network betweenness}.
\newblock \bibinfo{journal}{\emph{Social networks}} \bibinfo{volume}{27},
  \bibinfo{number}{1} (\bibinfo{year}{2005}), \bibinfo{pages}{31--38}.
\newblock


\bibitem[\protect\citeauthoryear{Fan, Liu, Lu, Xiu, and Chen}{Fan
  et~al\mbox{.}}{2017}]%
        {fan2017efficient}
\bibfield{author}{\bibinfo{person}{Changjun Fan}, \bibinfo{person}{Zhong Liu},
  \bibinfo{person}{Xin Lu}, \bibinfo{person}{Baoxin Xiu}, {and}
  \bibinfo{person}{Qing Chen}.} \bibinfo{year}{2017}\natexlab{}.
\newblock \showarticletitle{An efficient link prediction index for complex
  military organization}.
\newblock \bibinfo{journal}{\emph{Physica A: Statistical Mechanics and its
  Applications}}  \bibinfo{volume}{469} (\bibinfo{year}{2017}),
  \bibinfo{pages}{572--587}.
\newblock


\bibitem[\protect\citeauthoryear{Fan, Xiao, Xiu, and Lv}{Fan
  et~al\mbox{.}}{2014}]%
        {fan2014fuzzy}
\bibfield{author}{\bibinfo{person}{Changjun Fan}, \bibinfo{person}{Kaiming
  Xiao}, \bibinfo{person}{Baoxin Xiu}, {and} \bibinfo{person}{Guodong Lv}.}
  \bibinfo{year}{2014}\natexlab{}.
\newblock \showarticletitle{A fuzzy clustering algorithm to detect criminals
  without prior information}. In \bibinfo{booktitle}{\emph{Proceedings of the
  2014 IEEE/ACM International Conference on Advances in Social Networks
  Analysis and Mining}}. IEEE Press, \bibinfo{pages}{238--243}.
\newblock


\bibitem[\protect\citeauthoryear{Grover and Leskovec}{Grover and
  Leskovec}{2016}]%
        {grover2016node2vec}
\bibfield{author}{\bibinfo{person}{Aditya Grover} {and} \bibinfo{person}{Jure
  Leskovec}.} \bibinfo{year}{2016}\natexlab{}.
\newblock \showarticletitle{node2vec: Scalable feature learning for networks}.
  In \bibinfo{booktitle}{\emph{KDD}}.
\newblock


\bibitem[\protect\citeauthoryear{Hamilton, Ying, and Leskovec}{Hamilton
  et~al\mbox{.}}{2017a}]%
        {hamilton2017inductive}
\bibfield{author}{\bibinfo{person}{Will Hamilton}, \bibinfo{person}{Zhitao
  Ying}, {and} \bibinfo{person}{Jure Leskovec}.}
  \bibinfo{year}{2017}\natexlab{a}.
\newblock \showarticletitle{Inductive representation learning on large graphs}.
  In \bibinfo{booktitle}{\emph{NIPS}}.
\newblock


\bibitem[\protect\citeauthoryear{Hamilton, Ying, and Leskovec}{Hamilton
  et~al\mbox{.}}{2017b}]%
        {hamilton2017representation}
\bibfield{author}{\bibinfo{person}{William~L Hamilton}, \bibinfo{person}{Rex
  Ying}, {and} \bibinfo{person}{Jure Leskovec}.}
  \bibinfo{year}{2017}\natexlab{b}.
\newblock \showarticletitle{Representation learning on graphs: Methods and
  applications}.
\newblock \bibinfo{journal}{\emph{arXiv}} (\bibinfo{year}{2017}).
\newblock


\bibitem[\protect\citeauthoryear{Holme and Kim}{Holme and Kim}{2002}]%
        {holme2002growing}
\bibfield{author}{\bibinfo{person}{Petter Holme} {and}
  \bibinfo{person}{Beom~Jun Kim}.} \bibinfo{year}{2002}\natexlab{}.
\newblock \showarticletitle{Growing scale-free networks with tunable
  clustering}.
\newblock \bibinfo{journal}{\emph{Physical review E}} \bibinfo{volume}{65},
  \bibinfo{number}{2} (\bibinfo{year}{2002}), \bibinfo{pages}{026107}.
\newblock


\bibitem[\protect\citeauthoryear{Holme, Kim, Yoon, and Han}{Holme
  et~al\mbox{.}}{2002}]%
        {holme2002attack}
\bibfield{author}{\bibinfo{person}{Petter Holme}, \bibinfo{person}{Beom~Jun
  Kim}, \bibinfo{person}{Chang~No Yoon}, {and} \bibinfo{person}{Seung~Kee
  Han}.} \bibinfo{year}{2002}\natexlab{}.
\newblock \showarticletitle{Attack vulnerability of complex networks}.
\newblock \bibinfo{journal}{\emph{Physical review E}} \bibinfo{volume}{65},
  \bibinfo{number}{5} (\bibinfo{year}{2002}), \bibinfo{pages}{056109}.
\newblock


\bibitem[\protect\citeauthoryear{Kipf and Welling}{Kipf and Welling}{2017}]%
        {kipf2016semi}
\bibfield{author}{\bibinfo{person}{Thomas~N Kipf} {and} \bibinfo{person}{Max
  Welling}.} \bibinfo{year}{2017}\natexlab{}.
\newblock \showarticletitle{Semi-supervised classification with graph
  convolutional networks}. In \bibinfo{booktitle}{\emph{ICLR}}.
\newblock


\bibitem[\protect\citeauthoryear{Kourtellis, Alahakoon, Simha, Iamnitchi, and
  Tripathi}{Kourtellis et~al\mbox{.}}{2013}]%
        {kourtellis2013identifying}
\bibfield{author}{\bibinfo{person}{Nicolas Kourtellis},
  \bibinfo{person}{Tharaka Alahakoon}, \bibinfo{person}{Ramanuja Simha},
  \bibinfo{person}{Adriana Iamnitchi}, {and} \bibinfo{person}{Rahul Tripathi}.}
  \bibinfo{year}{2013}\natexlab{}.
\newblock \showarticletitle{Identifying high betweenness centrality nodes in
  large social networks}.
\newblock \bibinfo{journal}{\emph{Social Network Analysis and Mining}}
  \bibinfo{volume}{3}, \bibinfo{number}{4} (\bibinfo{year}{2013}).
\newblock


\bibitem[\protect\citeauthoryear{Kumbhare, Frincu, Raghavendra, and
  Prasanna}{Kumbhare et~al\mbox{.}}{2014}]%
        {kumbhare2014efficient}
\bibfield{author}{\bibinfo{person}{Alok~Gautam Kumbhare}, \bibinfo{person}{Marc
  Frincu}, \bibinfo{person}{Cauligi~S Raghavendra}, {and}
  \bibinfo{person}{Viktor~K Prasanna}.} \bibinfo{year}{2014}\natexlab{}.
\newblock \showarticletitle{Efficient extraction of high centrality vertices in
  distributed graphs}. In \bibinfo{booktitle}{\emph{HPEC}}. IEEE,
  \bibinfo{pages}{1--7}.
\newblock


\bibitem[\protect\citeauthoryear{Li, Han, and Wu}{Li et~al\mbox{.}}{2018}]%
        {li2018deeper}
\bibfield{author}{\bibinfo{person}{Qimai Li}, \bibinfo{person}{Zhichao Han},
  {and} \bibinfo{person}{Xiao-Ming Wu}.} \bibinfo{year}{2018}\natexlab{}.
\newblock \showarticletitle{Deeper Insights into Graph Convolutional Networks
  for Semi-Supervised Learning}.
\newblock \bibinfo{journal}{\emph{arXiv}} (\bibinfo{year}{2018}).
\newblock


\bibitem[\protect\citeauthoryear{Li, Tarlow, Brockschmidt, and Zemel}{Li
  et~al\mbox{.}}{2016}]%
        {li2015gated}
\bibfield{author}{\bibinfo{person}{Yujia Li}, \bibinfo{person}{Daniel Tarlow},
  \bibinfo{person}{Marc Brockschmidt}, {and} \bibinfo{person}{Richard Zemel}.}
  \bibinfo{year}{2016}\natexlab{}.
\newblock \showarticletitle{Gated graph sequence neural networks}. In
  \bibinfo{booktitle}{\emph{ICLR}}.
\newblock


\bibitem[\protect\citeauthoryear{Mahmoody, Tsourakakis, and Upfal}{Mahmoody
  et~al\mbox{.}}{2016}]%
        {mahmoody2016scalable}
\bibfield{author}{\bibinfo{person}{Ahmad Mahmoody},
  \bibinfo{person}{Charalampos~E Tsourakakis}, {and} \bibinfo{person}{Eli
  Upfal}.} \bibinfo{year}{2016}\natexlab{}.
\newblock \showarticletitle{Scalable betweenness centrality maximization via
  sampling}. In \bibinfo{booktitle}{\emph{KDD}}.
\newblock


\bibitem[\protect\citeauthoryear{Mahyar, Hasheminezhad, Ghalebi, Nazemian,
  Grosu, Movaghar, and Rabiee}{Mahyar et~al\mbox{.}}{2018}]%
        {mahyar2018compressive}
\bibfield{author}{\bibinfo{person}{Hamidreza Mahyar}, \bibinfo{person}{Rouzbeh
  Hasheminezhad}, \bibinfo{person}{Elahe Ghalebi}, \bibinfo{person}{Ali
  Nazemian}, \bibinfo{person}{Radu Grosu}, \bibinfo{person}{Ali Movaghar},
  {and} \bibinfo{person}{Hamid~R Rabiee}.} \bibinfo{year}{2018}\natexlab{}.
\newblock \showarticletitle{Compressive sensing of high betweenness centrality
  nodes in networks}.
\newblock \bibinfo{journal}{\emph{Physica A: Statistical Mechanics and its
  Applications}}  \bibinfo{volume}{497} (\bibinfo{year}{2018}),
  \bibinfo{pages}{166--184}.
\newblock


\bibitem[\protect\citeauthoryear{Newman}{Newman}{2006}]%
        {newman2006modularity}
\bibfield{author}{\bibinfo{person}{Mark~EJ Newman}.}
  \bibinfo{year}{2006}\natexlab{}.
\newblock \showarticletitle{Modularity and community structure in networks}.
\newblock \bibinfo{journal}{\emph{PNAS}} \bibinfo{volume}{103},
  \bibinfo{number}{23} (\bibinfo{year}{2006}).
\newblock


\bibitem[\protect\citeauthoryear{Peixoto}{Peixoto}{2014}]%
        {peixoto2014graph}
\bibfield{author}{\bibinfo{person}{Tiago~P Peixoto}.}
  \bibinfo{year}{2014}\natexlab{}.
\newblock \showarticletitle{The graph-tool python library}.
\newblock \bibinfo{journal}{\emph{figshare}} (\bibinfo{year}{2014}).
\newblock


\bibitem[\protect\citeauthoryear{Perozzi, Al-Rfou, and Skiena}{Perozzi
  et~al\mbox{.}}{2014}]%
        {perozzi2014deepwalk}
\bibfield{author}{\bibinfo{person}{Bryan Perozzi}, \bibinfo{person}{Rami
  Al-Rfou}, {and} \bibinfo{person}{Steven Skiena}.}
  \bibinfo{year}{2014}\natexlab{}.
\newblock \showarticletitle{Deepwalk: Online learning of social
  representations}. In \bibinfo{booktitle}{\emph{KDD}}.
\newblock


\bibitem[\protect\citeauthoryear{Pfeffer and Carley}{Pfeffer and
  Carley}{2012}]%
        {pfeffer2012k}
\bibfield{author}{\bibinfo{person}{J{\"u}rgen Pfeffer} {and}
  \bibinfo{person}{Kathleen~M Carley}.} \bibinfo{year}{2012}\natexlab{}.
\newblock \showarticletitle{k-centralities: Local approximations of global
  measures based on shortest paths}. In \bibinfo{booktitle}{\emph{WWW}}.
\newblock


\bibitem[\protect\citeauthoryear{Riondato and Kornaropoulos}{Riondato and
  Kornaropoulos}{2016}]%
        {riondato2016fast}
\bibfield{author}{\bibinfo{person}{Matteo Riondato} {and}
  \bibinfo{person}{Evgenios~M Kornaropoulos}.} \bibinfo{year}{2016}\natexlab{}.
\newblock \showarticletitle{Fast approximation of betweenness centrality
  through sampling}.
\newblock \bibinfo{journal}{\emph{DMKD}} \bibinfo{volume}{30},
  \bibinfo{number}{2} (\bibinfo{year}{2016}).
\newblock


\bibitem[\protect\citeauthoryear{Riondato and Upfal}{Riondato and
  Upfal}{2018}]%
        {riondato2018abra}
\bibfield{author}{\bibinfo{person}{Matteo Riondato} {and} \bibinfo{person}{Eli
  Upfal}.} \bibinfo{year}{2018}\natexlab{}.
\newblock \showarticletitle{ABRA: Approximating betweenness centrality in
  static and dynamic graphs with rademacher averages}.
\newblock \bibinfo{journal}{\emph{TKDD}} \bibinfo{volume}{12},
  \bibinfo{number}{5} (\bibinfo{year}{2018}).
\newblock


\bibitem[\protect\citeauthoryear{Staudt, Sazonovs, and Meyerhenke}{Staudt
  et~al\mbox{.}}{2016}]%
        {staudt2016networkit}
\bibfield{author}{\bibinfo{person}{Christian~L Staudt},
  \bibinfo{person}{Aleksejs Sazonovs}, {and} \bibinfo{person}{Henning
  Meyerhenke}.} \bibinfo{year}{2016}\natexlab{}.
\newblock \showarticletitle{NetworKit: A tool suite for large-scale complex
  network analysis}.
\newblock \bibinfo{journal}{\emph{Network Science}} \bibinfo{volume}{4},
  \bibinfo{number}{4} (\bibinfo{year}{2016}), \bibinfo{pages}{508--530}.
\newblock


\bibitem[\protect\citeauthoryear{Tang, Qu, Wang, Zhang, Yan, and Mei}{Tang
  et~al\mbox{.}}{2015}]%
        {tang2015line}
\bibfield{author}{\bibinfo{person}{Jian Tang}, \bibinfo{person}{Meng Qu},
  \bibinfo{person}{Mingzhe Wang}, \bibinfo{person}{Ming Zhang},
  \bibinfo{person}{Jun Yan}, {and} \bibinfo{person}{Qiaozhu Mei}.}
  \bibinfo{year}{2015}\natexlab{}.
\newblock \showarticletitle{Line: Large-scale information network embedding}.
  In \bibinfo{booktitle}{\emph{WWW}}.
\newblock


\bibitem[\protect\citeauthoryear{Velickovic, Cucurull, Casanova, Romero, Lio,
  and Bengio}{Velickovic et~al\mbox{.}}{2018}]%
        {velickovic2017graph}
\bibfield{author}{\bibinfo{person}{Petar Velickovic}, \bibinfo{person}{Guillem
  Cucurull}, \bibinfo{person}{Arantxa Casanova}, \bibinfo{person}{Adriana
  Romero}, \bibinfo{person}{Pietro Lio}, {and} \bibinfo{person}{Yoshua
  Bengio}.} \bibinfo{year}{2018}\natexlab{}.
\newblock \showarticletitle{Graph attention networks}. In
  \bibinfo{booktitle}{\emph{ICLR}}.
\newblock


\bibitem[\protect\citeauthoryear{Xu, Li, Tian, Sonobe, Kawarabayashi, and
  Jegelka}{Xu et~al\mbox{.}}{2018}]%
        {xu2018representation}
\bibfield{author}{\bibinfo{person}{Keyulu Xu}, \bibinfo{person}{Chengtao Li},
  \bibinfo{person}{Yonglong Tian}, \bibinfo{person}{Tomohiro Sonobe},
  \bibinfo{person}{Ken-ichi Kawarabayashi}, {and} \bibinfo{person}{Stefanie
  Jegelka}.} \bibinfo{year}{2018}\natexlab{}.
\newblock \showarticletitle{Representation Learning on Graphs with Jumping
  Knowledge Networks}. In \bibinfo{booktitle}{\emph{ICML}}.
\newblock


\bibitem[\protect\citeauthoryear{Yoshida}{Yoshida}{2014}]%
        {yoshida2014almost}
\bibfield{author}{\bibinfo{person}{Yuichi Yoshida}.}
  \bibinfo{year}{2014}\natexlab{}.
\newblock \showarticletitle{Almost linear-time algorithms for adaptive
  betweenness centrality using hypergraph sketches}. In
  \bibinfo{booktitle}{\emph{KDD}}.
\newblock


\end{thebibliography}
% \clearpage
% \input{Section/appendix}
\end{document}